%% file: main.tex
\documentclass[sigconf,authorversion]{acmart}
\AtBeginDocument{%
  \providecommand\BibTeX{{%
    \normalfont B\kern-0.5em{\scshape i\kern-0.25em b}\kern-0.8em\TeX}}}

\setcopyright{acmcopyright}
\copyrightyear{2025}
\acmYear{2025}
\setcopyright{acmlicensed}\acmConference[CHI '25]{CHI Conference on Human
Factors in Computing Systems}{April 26-May 1, 2025}{Yokohama, Japan}
\acmBooktitle{CHI Conference on Human Factors in Computing Systems (CHI
'25), April 26-May 1, 2025, Yokohama, Japan}
\acmDOI{10.1145/3706598.3714087}
\acmISBN{979-8-4007-1394-1/25/04}

\usepackage{xspace}
\newcommand{\ie}{i.e.{\xspace}}
\newcommand{\eg}{e.g.{\xspace}}
\newcommand{\vs}{vs.{\xspace}}
\newcommand{\rev}[1]{#1}
\newcommand{\qt}[1]{\textit{``#1''}}
\newcommand{\pqt}[2]{\textit{``#1''}{\,}{\small-#2}}

\newcommand{\etal}{{\xspace}et~al.{\xspace}}

\newcommand{\anova}[4]{{$F_{#1,#2}=#3$, $p<#4$}}
\newcommand{\anovans}[3]{{$F_{#1,#2}=#3$, $p>.05$}}

\usepackage{amsmath}
\usepackage{graphicx}
\usepackage{wrapfig}
\usepackage{subcaption}
\usepackage{multirow}
\begin{document}
\newcommand{\xinyu}[1]{\authorcomment{RED}{Xinyu}{#1}}
\newcommand{\jian}[1]{\authorcomment{BLUE}{JZ}{#1}}
\newcommand{\ryan}[1]{\authorcomment{GREEN}{Ryan}{#1}}


\raggedbottom
\newcommand{\sysname}{Brickify}
\title[\textsc{\sysname{}}]{\textsc{\sysname{}}: Enabling Expressive Design Intent Specification through Direct Manipulation on Design Tokens}

\author{Xinyu Shi}
\affiliation{%
  \institution{School of Computer Science \\ University of Waterloo}
  \city{Waterloo}
  \state{ON}
  \country{Canada}
}
\email{xinyu.shi@uwaterloo.ca}

\author{Yinghou Wang}
\affiliation{%
  \institution{Graduate School of Design \\ Harvard University}
  \city{Cambridge}
  \state{MA}
  \country{United States}
}
\email{yinghouwang@gsd.harvard.edu}

\author{Ryan Rossi}
\affiliation{%
  \institution{Adobe Research}
  \city{San Jose}
  \state{CA}
  \country{United States}
}
\email{ryrossi@adobe.com}

\author{Jian Zhao}
\affiliation{%
  \institution{School of Computer Science \\ University of Waterloo}
  \city{Waterloo}
  \state{ON}
  \country{Canada}
}
\email{jianzhao@uwaterloo.ca}


\begin{abstract}

Expressing design intent using natural language prompts requires designers to verbalize the ambiguous visual details concisely, which can be challenging or even impossible. 
To address this, we introduce \textsc{\sysname{}}, a \emph{visual-centric} interaction paradigm --- expressing design intent through \emph{direct manipulation on design tokens}.
\textsc{\sysname{}} extracts visual elements (\eg, subject, style, and color) from reference images and converts them into interactive and reusable design tokens that can be directly manipulated (\eg, resize, group, link, etc.) to form the \emph{visual lexicon}. 
The lexicon reflects users' intent for both \emph{what} visual elements are desired and \emph{how} to construct them into a whole.
We developed \sysname{} to demonstrate how AI models can interpret and execute the \emph{visual lexicon} through an end-to-end pipeline.
In a user study, experienced designers found \textsc{\sysname{}} more efficient and intuitive than text-based prompts, allowing them to describe visual details, explore alternatives, and refine complex designs with greater ease and control.


\end{abstract}

\begin{teaserfigure}
    \centering
    \vspace{-1mm}
    \includegraphics[width=1.\linewidth]{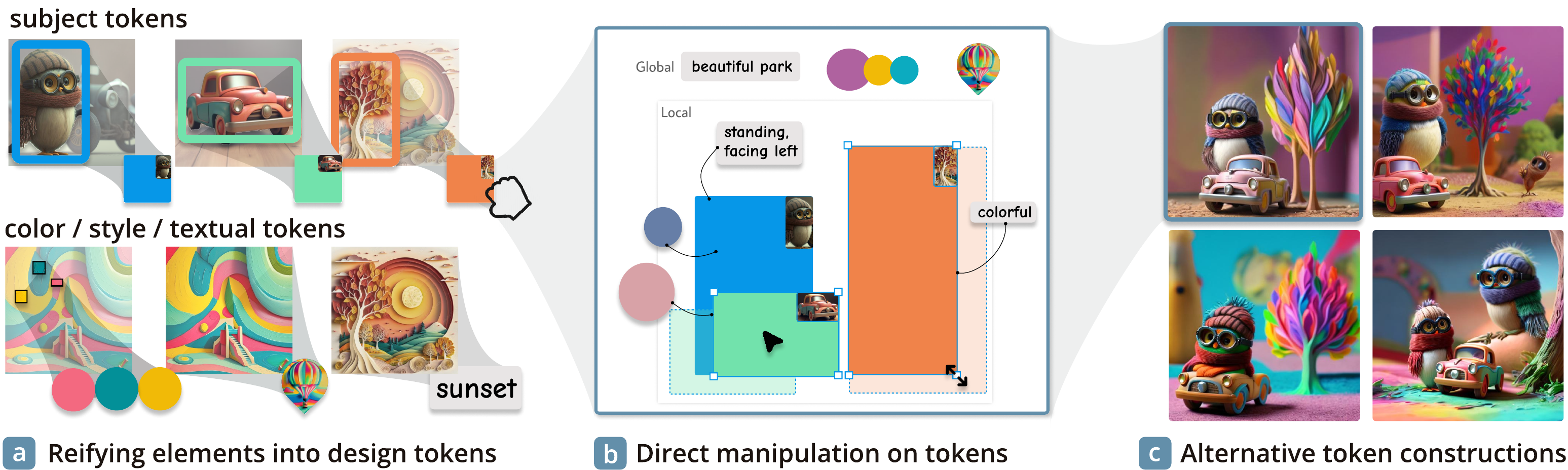}
    \vspace{-5mm}
    \caption{\textsc{\sysname{}} introduces a visual-centric interaction paradigm to specify design intent for controllable image generation: (a) users start with reference images and identify design elements (subjects, styles, colors, and concepts) which can be reified into interactive, reusable design tokens; (b) then directly manipulate on these tokens to build a visual lexicon to express how to construct these elements as a whole; (c) explore alternative compositions by reusing tokens and refining the visual lexicon.
    }
    \label{fig:teaser}
    \Description{The figure illustrates the \textsc{\sysname{}} interaction paradigm in three stages: (a) Reifying elements into design tokens: users begin by selecting specific subjects, styles, colors, and textual concepts from reference images. These elements are transformed into interactive design tokens. For example, subject tokens (such as an owl or a car) and style tokens (like a colorful abstract pattern) are extracted from the images. (b) Direct manipulation on tokens: users can manipulate these design tokens to construct a visual lexicon, specifying relationships between elements. The figure shows the interface for adjusting positions, sizes, and linking textual descriptors such as ``colorful'' or ``standing, facing left.'' (c) Alternative token constructions: Users can explore variations by reusing design tokens and fine-tuning the visual lexicon. The figure shows examples of different compositions generated by rearranging and modifying the visual lexicon, resulting in alternative creative outcomes.}
\end{teaserfigure}




\maketitle

\input{texfiles/1-introduction.tex}
\input{texfiles/2-relatedwork.tex}

\input{texfiles/3-method.tex}

\input{texfiles/4-duplo}
\input{texfiles/5-usage-scenario.tex}

\input{texfiles/6-implementation.tex}

\input{texfiles/7-study-design.tex}

\input{texfiles/8-study-results}

\input{texfiles/9-discussion.tex}

\input{texfiles/10-conclusion.tex}

\begin{acks}
We sincerely thank Xueguang Ma, Ryan Yen,  and anonymous reviewers for their insightful suggestions that have led to a great improvement of this work. 
We also extend our appreciation to Annie Sun and Hakeerat Singh Mayall for their kind help on a few early-stage prototypes.
We would also like to thank all our participants for their time and valuable input. 
This work is supported in part by the Discovery Grant (RGPIN-2020-03966) from the NSERC (Natural Sciences and Engineering Research Council of Canada) and a gift fund from Adobe Systems Inc.

We acknowledge that much of our work takes place on the traditional territory of the Neutral, Anishinaabeg, and Haudenosaunee peoples. 
Our main campus is located on the Haldimand Tract, the land granted to the Six Nations that includes six miles on each side of the Grand River.
\end{acks}

\bibliographystyle{ACM-Reference-Format}
\bibliography{references.bib}

\appendix
\input{texfiles/12-table}
\newpage
\input{texfiles/11-appendix.tex}

\end{document}

%% file: texfiles/1-introduction.tex
\section{Introduction}


Design is about the choices of visual elements (\eg, subject, style, color) and the construction of them towards an intended effect~\cite{dondis1974primer}; the decisions involved in this process are the art of design, embraced with a designer's creativity~\cite{goldschmidt2005good}. 
However, current text-to-image generation tools (\eg, DALL-E, Midjourney) shift design decision-making from designers to models, which are designed to \emph{create for users} rather than \emph{work with designers}~\cite{Satyanarayan2024Intelligence, weisz2024design, subramonyam2024bridging}. 
Despite their ability to produce aesthetically appealing images for casual use, without designer's thoughtful planning, they lack the capability to create meaningful and professional design solutions that effectively convey intended visual messages~\cite{shi2023understanding, chen2023next, choi2023creativeconnect}.
One key barrier to keeping designers \emph{in-the-loop} is communication~\cite{weisz2024design, subramonyam2024bridging} --- generative AI tools are designed to receive instructions \emph{textually}, while designers often prefer to think and communicate \emph{visually}~\cite{tschimmel2012design, ware2010visual}. 

\begin{figure}[tbp]
    \vspace{7mm}
    \includegraphics[width=\linewidth]{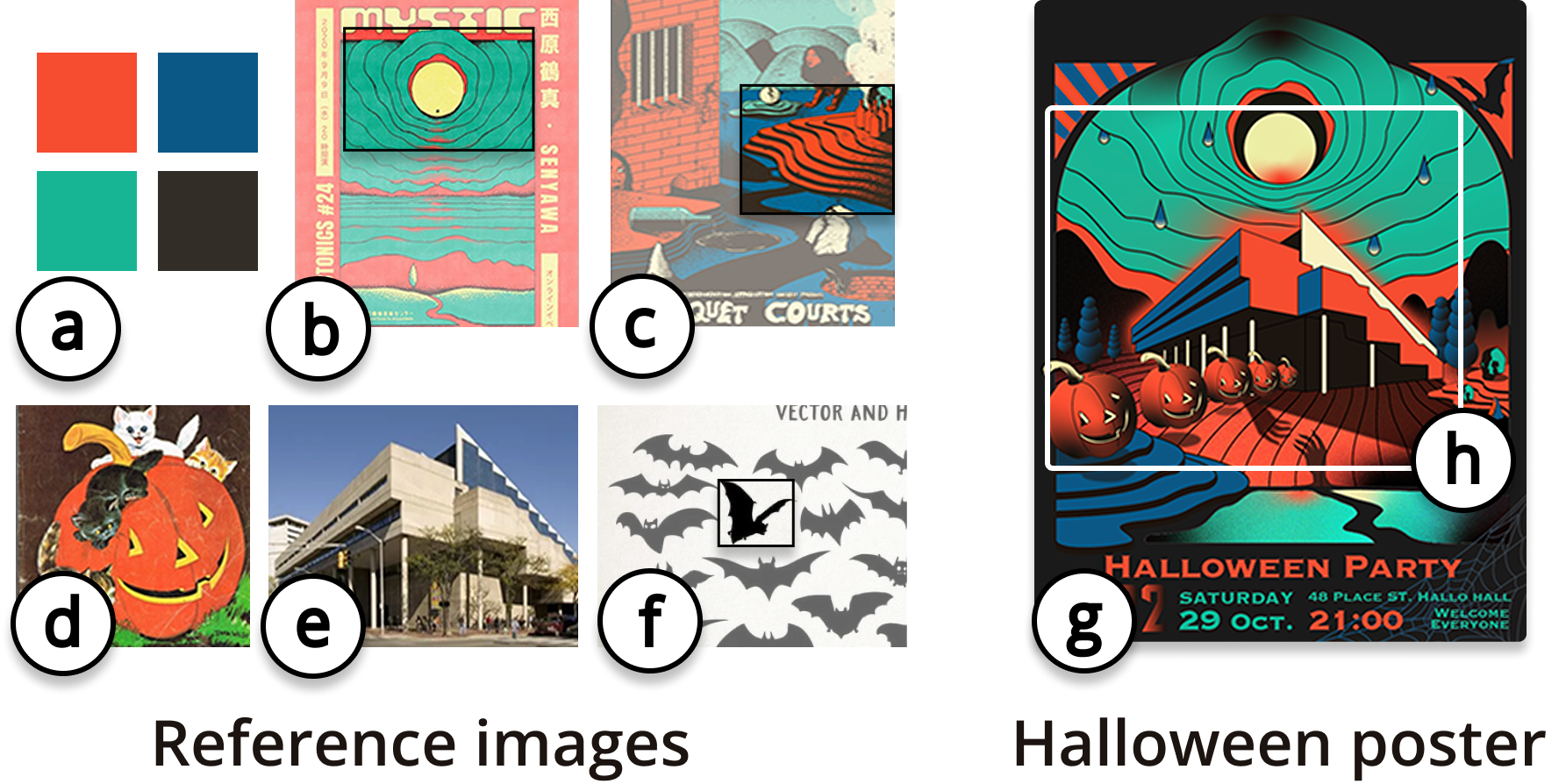}
    \caption{Design example of a Halloween party poster, showing (a) the color palette, (b-f) reference images with highlighted elements, and (g) the envisioned poster in designer's mind. (h) illustrates the spatial relationships between pumpkins, building, and moon. We have obtained the designer's consent to include this design in the paper.}
    \label{fig:halloween}
    \Description{The figure illustrates how different reference images contribute to the composition of a Halloween poster. On the left, various reference images are shown: a color palette with four swatches (red, blue, green, and black), a stylized sun over water, an architectural design highlighting a red curved shape, a classic Halloween pumpkin decoration, a modern building with a diagonal roof structure, and a vector-style bat illustration with one bat highlighted. On the right, the Halloween poster incorporates elements from these references. The figure demonstrates how designers extract and recombine elements from multiple sources to create a cohesive design.}
\end{figure}

Consider the design case of crafting a Halloween poster: 
the designer collects some reference images (Figure~\ref{fig:halloween}a-\ref{fig:halloween}f), thinks about what visual elements to use, and after a while, forms a rough idea (Figure~\ref{fig:halloween}g) about how to compose them into a whole, and wants to work with generative AI tools for quick prototyping.
However, the following three challenges arise.

First, clearly describing \emph{what} visual elements to use in natural language is challenging due to inconsistent naming standards~\cite{heider1972universals, murphy2004big}.
For example, colors in Figure~\ref{fig:halloween}a might be described as \textit{``light red, greenish teal, navy blue, with greyish black''} by some designers, and simply as \textit{``orange, green, blue, and black''} by others. 
These subtle differences can lead to significant variations in hues and shades.
Uploading reference images for model to ``see''~\cite{choi2023creativeconnect, peng2024designprompt} can help when the desired element dominates the image as in Figure~\ref{fig:halloween}d. 
However, in complex images as Figure~\ref{fig:halloween}c, specifying an exact element is harder. 
Designers might describe it as \textit{``the abstract shape with curves in the middle right''}, but models often struggle with such ambiguous descriptions regarding the shape and position.

Further, precisely verbalizing \emph{how} to construct those visual elements into a whole is hard. 
Deciding on relative scale and proximity is to relate the isolated elements with each other as interacting parts~\cite{dondis1974primer, ware2010visual}.
However, both scale and proximity are continuous values, while language is often too discrete for fine-grained instructions.
For example, describing Figure~\ref{fig:halloween}h as \textit{``five pumpkins in front of a large building, with a moon above''} lacks precise size and position details, making it hard for the model to interpret their nuanced spatial relationship.
The designer might finally obtain a reasonable version after multiple rounds of conversations with the tool, \eg, \textit{``make the building a little bit larger''}, but this process is often time-consuming and tedious without guaranteed outcomes.


Lastly, the choices of elemental construction are infinite; however, prompting with texts limits the flexibility to \emph{reuse} those visual elements to explore alternative constructions. 
Designers need to \emph{copy-and-paste} an entire paragraph from previous prompts, then modify certain parts to change relationships or replace visual elements.
Each iteration of exploration requires users to manually track where to change and where to keep among lengthy texts, which is tedious and error-prone.
It is because existing generative AI tools treat each prompt as an independent request, without a mechanism to selectively separate persistent information (\eg, visual characteristics) from single-use prompts, making it inefficient to share information across multi-turn interactions.

To better align generative AI tools with designers' visual thinking process, we need a \emph{visual-centric} interaction paradigm with versatile expressive power \rev{to facilitate graphic design that builds upon visual assets.}
We propose \textsc{\sysname{}} --- specifying design intent via \emph{interactive design tokens} that clearly carry the information of \emph{what} primitive design elements to use (\eg, subject, color, style). 
Design tokens can be directly manipulated (\eg, drag-and-drop, resize, move, group, and link) to allow designers to precisely plan \emph{how} the visual elements are constructed. 
The resultant construction --- \emph{visual lexicon} --- can be translated into control signals for AI to faithfully generate the desired outcome, \eg, spatial layouts, relative scales, and the effect radius (\eg, applying a color to a specific subject or the entire image).
Since design tokens can be persistent throughout the design process, users can efficiently \emph{reuse} and recombine elements, avoiding the need to start from scratch for each iteration.
We implemented and iteratively refined this interaction paradigm of \textsc{\sysname{}} in an interactive system named \sysname{} (\textsc{\sysname{}} refers to the paradigm while \sysname{} denotes the system).

We evaluated the interaction paradigm \textsc{\sysname{}} and its implementation in \sysname{} through an in-lab user study with 12 experienced designers. 
In a replication task simulating a scenario where designers have a clear intent, we compared \textsc{\sysname{}} with \emph{textual-centric prompting}, finding that participants expressed design intent more precisely with less cognitive effort using \textsc{\sysname{}} and performed refinements faster, especially in complex designs. 
In an open-ended task with a simulated exploratory design scenario, designers found \sysname{} provided a controllable, expressive, and engaging design experience.
Our findings offer insights for future research on designing interaction mediums for human-AI co-creation in broader visual design contexts.

%% file: texfiles/2-relatedwork.tex
\section{Related Work}
AI-assisted graphic design involves \emph{forming}, \emph{specifying}, and \emph{realizing} design intentions. 
We review how mood boards help form intentions, explore interactive techniques for specifying them, and summarize personalized image generation approaches to translate intentions into design outcomes.


\vspace{-1mm}
\subsection{Forming Intent from Reference Imagery}
A common approach to develop design ideas is using references.
Designers often start with exploring and organizing inspirations, then identify key design elements and strategically compose them~\cite{finke1996creative}.

\textbf{Mood board usage.}
To support divergent thinking, designers often start with creating the mood board \cite{eckert2000sources}, a collection of images, shapes, colors, and other visual stimuli, as an aid to conduct visual research --- framing, aligning, paradoxing, abstracting, and directing the design~\cite{lucero2012framing}. 
Mood boards are intentionally ambiguous, allowing different interpretations and serving as a tool for exploring creative possibilities~\cite{finke1996creative}. 
How designers organize mood boards reflects their intended use, with studies exploring image arrangement in digital drawing and other fields~\cite{holinaty2021supporting, endrissat2016visual, lucero2005mood}. 
Recent works~\cite{koch2019may, koch2020imagesense, koch2020semanticcollage} enhance mood boards with AI for semantic clustering, image recommendations, and material arrangement. 
Designers often decompose images into sub-elements~\cite{dondis1974primer} and then integrate them into cohesive designs~\cite{garner2001problem}. 
Tools like MoodCubes~\cite{ivanov2022moodcubes} and MetaMap~\cite{kang2021metamap} help break down elements and enhance mood boards.
Current generative AI tools allow users to upload reference images but lack flexibility in arranging or specifying sub-elements like on a mood board. Users cannot easily select which parts to use. Our approach integrates mood boards, enabling users to specify \emph{what} sub-elements to use by converting them into design tokens.


\textbf{Recombination of recognized elements.}
After identifying the elements of interest on the mood board, researchers have created a variety of tools to support the elements recombination. 
VisiBlends~\cite{chilton2019visiblends} and VisiFit~\cite{chilton2021visifit} enable the blending of two semantic objects.
Building on this, PopBlends~\cite{wang2023popblends} explores strategies for merging two conceptual keywords into pop culture images using large language models.
Tools like 3DALL-E~\cite{liu20233dall} and CreativeConnect~\cite{choi2023creativeconnect} propose workflows for generating 3D designs and graphic sketches by suggesting keywords from design references and then refining them into detailed text prompts.
Beyond keywords, Artinter~\cite{chung2023artinter} and GANCollage~\cite{wan2023gancollage} enhance designer-client communication and style mixing by combining keywords and example imagery.
Existing tools typically emphasize recombination outcomes under predefined rules for task-specific usage, such as blending by shape contours or mixing by styles. 
However, users have limited control over \emph{how} to combine. 
This paper addresses this issue by enabling users to expressively construct the relationships among elements through direct manipulation of reified interactive tokens.

\vspace{-1mm}
\subsection{Specifying Intent by Interactive Strategies}
\rev{
The challenge of specifying design intent solely with natural language is well recognized \cite{weisz2024design, chen2023next, simkute2024ironies, tankelevitch2024metacognitive}. 
Subramonyam~\etal~\cite{subramonyam2024bridging} theorize how users translate their goals into clear intentions, highlighting the \emph{instruction gap} where generative models are highly sensitive to language precision but human language tolerates variants in expression to communicate similar meaning.
This challenge is evident across studies in various design domains, where casual users~\cite{mahdavi2024ai}, graphic designers~\cite{kulkarni2023word}, manufacturing designers~\cite{gmeiner2023exploring}, and game professionals~\cite{vimpari2023adapt} struggle to articulate visual intent through the tedious process of prompt engineering.
This friction of translating \emph{visual} design into a \emph{verbal} medium can be understood through the lens of design methodology literature~\cite{ward1984design, cross1990nature, tomes1998talking}.
}
Recent efforts to facilitate the intent expression fall into three categories: 1) decomposing the lengthy prompt into modular ones; 2) augmenting the text prompt with other modalities; and 3) resolving ambiguities in text prompts through direct manipulation \cite{shneiderman1983direct}.


\textbf{Modular prompting.}
Managing lengthy prompts is challenging, and many research has explored to modularize them into manageable pieces.
For instance, AI-chains~\cite{wu2022ai} enables chaining individual prompts for end-to-end execution. 
In the similar spirit, ChainForge~\cite{arawjo2024chainforge} supports comparing prompt variations between models.
In visual design, Keyframer~\cite{keyframer} uses ``decomposed prompting'' for step-by-step animation design. 
Similarly, Spellburst~\cite{angert2023spellburst} and ComfyUI~\cite{ComfyUI} employ node-based interfaces to modularize creative coding and image generation, integrating diverse control signals.
Although these tools aid in flexibly writing and modifying prompts and allow viewing intermediate results, they still require users to specify their visual intentions in text, which does not fully address the precision of intent specification.

\textbf{Multimodal prompting.}
\rev{Most recent commercialized tools (\eg, DALL$\cdot$E3 \cite{Dalle3}, Adobe Firefly \cite{Firefly}, MidJourney \cite{MidJourney}, Flux AI \cite{FluxAI}) allow users upload images as global content and/or style references; however, users cannot add local annotations on the image to further illustrate their fine-grained intentions.
Kaiber Superstudio \cite{Kaiber} supports character-consistent generation by allowing users to specify a single local subject but does not support multiple subjects. Krea.ai \cite{Krea} enables users to train on their own assets but requires multiple instances for each subject.
}
Research efforts have also investigated different strategies for multimodal prompting.
For example, DesignPrompt \cite{peng2024designprompt} allows users to input texts, images, and colors, then help translate them into a final textual prompt. 
However, translating visuals into text often leads to information loss, making it difficult to capture precise visual identities.
PromptCharm \cite{wang2024promptcharm} enables users to examine which part of the generated image corresponds to which part of the text prompt.
\rev{Sarukkai~\etal~\cite{Sarukkai2024block} introduce the coarse-to-fine sketch guided image generation.}
Although these tools are helpful in clarifying \emph{what} visual elements to use to some degree, they fail to help users express \emph{how} multiple visual elements relate with each other because they typically treat each input modality in isolation.

\textbf{Prompting through direct manipulation.}
Some tools employ visual metaphors to help users specify their intentions through directly manipulating on visual objects rather than text prompts. 
For instance, TaleBrush \cite{chung2022talebrush} uses sketch lines to indicate narrative transitions, and PromptPaint \cite{chung2023promptpaint} offers a paint-like interface for semantic prompt interpolations. 
Gestures are utilized to represent editing intentions, for example, drawing masks to inpaint \cite{avrahami2023blended}, adding colored strokes to recolorize \cite{zhang2017real}, and dragging points to edit pose or facial expressions \cite{pan2023drag}.
While intuitive, these methods are task-specific, requiring users to re-learn interactions for each use case.
Recent work extends this direction by aiming to integrate graphical user interfaces (GUI) with natural language interfaces (NLI). 
DirectGPT \cite{masson2023directgpt} allows users to drag and drop graphical elements onto prompts, but these elements act as isolated symbols, losing their spatial relationships. 
DynaVis \cite{vaithilingam2024dynavis} combines NLIs with dynamically generated GUI widgets for visualization authoring, but requires users to specify the edit intention in text. 
Both DirectGPT and DynaVis address the ambiguity of continuous numerical expressions in textual prompts, yet they still force users to think textually and work primarily on texts.
\subsection{Translating Intent into Design Outcomes through Computer Vision Techniques}
Text-to-image generation is an active research topic since diffusion models emerging~\cite{ho2020denoising, song2019generative}, 
there are many amazing work has made the performance of text-to-image generation has reached a height that never been reached,
such as DALL-E~\cite{ramesh2021zero, ramesh2022hierarchical, betker2023improving}, GLIDE~\cite{nichol2021glide}, Stable Diffusion~\cite{rombach2022high}, Imagen~\cite{saharia2022photorealistic}, etc.
Research has also focused on improving controllability in image generation and editing.


\textbf{Subject-driven and style-specific personalized generation.}
The task of \emph{personalization}, introduced by Cohen~\etal~\cite{cohen2022my}, aims to incorporate user-provided concepts absent from the training data into generated results.
Early methods like Textual-Inversion~\cite{gal2022image} and DreamBooth~\cite{ruiz2023dreambooth} learn a \emph{single subject} from \emph{several user images}, while Custom Diffusion~\cite{kumari2022customdiffusion} and SVDiff~\cite{han2023svdiff} extend this to \emph{multiple subjects} for recombination.
Later, ELITE~\cite{wei2023elite} and E4T-diffusion~\cite{gal2023encoder} make it possible to learn a \emph{single subject} from \emph{one image} but require the subject to be visually dominant.
Break-A-Scene~\cite{avrahami2023break} allows learning \emph{multiple subjects} from a \emph{single image} with loose segmentation masks. 
Personalization also applies to styles: StyleDrop~\cite{sohn2023styledrop} fine-tunes models for style customization, while Style-Align~\cite{hertz2023style} ensures style consistency without fine-tuning.

\textbf{Spatial-aware controllable image generation and editing.}
To guide the large-scale pre-trained diffusion models to generate images following the spatially-localized conditions (\eg, depth map, segmentation, pose, etc.), Controlnet~\cite{zhang2023adding} embeds task-specific networks, while T2I-Adapter~\cite{mou2024t2i} offers a lighter adapter-based solution, and Composer~\cite{huang2023composer} provides more flexible control with a larger model.
For local image editing, initial methods~\cite{avrahami2022blended, hertz2022prompt, bar2022text2live} rely on precise text descriptions. 
Imagic~\cite{kawar2023imagic} allows description-free editing but requires time-intensive fine-tuning, whereas Blended Latent Diffusion~\cite{avrahami2023blended} enables faster editing without fine-tuning.

The advancements in the computer vision field are powerful and hold great potential. 
However, without intuitive interactions that allow users to precisely express their underlying intent, users cannot fully benefit from these capabilities. 
Our work focuses on innovating the user interaction, using these off-the-shelf methods as the technical foundation to realize our proposed \emph{visual-centric} interaction paradigm.



%% file: texfiles/3-method.tex
\section{Iterative User-Centered Design}

In collaboration with designers for about nine months, we employed an iterative design approach to define the \emph{visual-centric} interaction paradigm, \textsc{\sysname{}}, and develop the \sysname{} system. 
The design process consists of four stages (noted as \textbf{S1-4}), with different participants involved in each stage.
In this section, we introduce the participants and procedures for S1, S2, and S3, and discuss the findings and derived design goals from S1 (Section~\ref{sec:design_s1}). 
To provide a holistic view, we briefly highlight the key design decisions made from S2 (Section~\ref{sec:design_s2}) and S3 (Section~\ref{sec:design_s3}), with further details provided in Section~\ref{sec:paradigm}.

\begin{itemize}
    \item[] \emph{S1: Problem understanding (2 months)} --- interviews with six designers to identify challenges in using generative AI tools.  
    \item[] \emph{S2: Early Prototyping (4 months)} --- weekly co-design sessions with an expert designer to design the interaction paradigm and develop the early working prototype;
    \item[] \emph{S3: Prototype Iteration (2 months)} --- informal testing involved six designers to collect feedback and iteratively refine the design;
    \item[] \emph{S4: System Evaluation (1 month)} ---  a user study with two tasks compares the visual-centric interaction paradigm of \textsc{\sysname{}} with the textual-centric one and examines how users interact with \sysname{}, which will be described in Section~\ref{sec:study_design} and Section~\ref{sec:study_result}.
\end{itemize}

\subsection{Problem Understanding: Interview Study}
\label{sec:design_s1}
We conducted semi-structured interviews with six design experts (E1-6), with the aim to understand: 
(1) how designers approach prompting the generative AI models to craft graphic designs; 
and (2) what challenges they have encountered.

\subsubsection{Participants and Procedure}
Participants were recruited via email lists and social media, screened through a pre-test survey on design experience and familiarity with text-to-image generation tools. 
All had over two years of experience in graphic design, familiar with and regularly used the text-to-image generation tools in their work.
Participants provided consent and were compensated with \$20 for a 45-minute study session. 
We asked each participant to share at least one recent design project involving using text-to-image AI tools to reflect on how they use them.


\subsubsection{Identified Challenges}
We summarize the following challenges designers encountered when using generative AI tools. 

\textbf{C1: Failure to convey designers' attended elements to AI.}
Participants consistently began their design projects with visual research, using mood boards to collect inspirational images on a canvas in Figma (5/6) or Photoshop (1/6). 
They emphasized \pqt{it's for understanding how pieces can fit together in my head.}{E3}, aligning with prior studies~\cite{koch2020imagesense, lucero2012framing, holinaty2021supporting}. 
Some participants (E1, E2, E4) grouped references spatially by element type, while others (E3, E5, E6) used annotations to mark elements.
E5 explained, \qt{I'm not looking at the whole image, just the parts that matter to my design.}
This selective focus, known as \emph{active vision}~\cite{findlay2003active, goodwin2015professional}, is central to \emph{visual thinking}~\cite{ware2010visual}. 
While designers use \emph{active vision} to pinpoint specific details such as a particular texture, a color theme, or the composition of shapes, AI models often lack the ability to recognize or prioritize these details in the same way. 
As E4 explained, \qt{The AI seems to understand the image globally, but I need it to work with specific parts.}
Designers have to communicate the visual elements they are focusing on with AI through natural language, a medium that \pqt{super hard for describing fine visual nuances.}{E1}

\textbf{C2: Difficulty in verbalizing element relationships.}
Participants emphasized the complexity of composing visual elements in a design, as E2 described, \qt{like constructing a house, you must place each brick properly.}
E3 explained, \qt{It's not just about having the pieces; how to balance their weight is also important --- some are more important, while others just for decorating.}
Designers must create focal points, balance hierarchy, and manage spatial placements, but articulating these relationships in words is difficult, especially when elements are intertwined. 
As E6 noted, \qt{planning that (how to compose them) in my head is hard enough, translating (the entire mental image) into a sentence feels much more difficult, I often sketch them down then describe.}
This challenge often forces designers to simplify their ideas.
As E2 mentioned, \qt{I tend to only ask for simple structures like centering a certain one (element)}, but such compromise often results in outputs that \qt{lack the balance and spatial nuance we have been taught and always pursue in design.}

\textbf{C3: Inefficiencies in iterative refinements.}
Designers face significant challenges in refining visuals through current text-to-image tools, as the conversations with AI is linear and lacks the mechanism to share key visual information in multi-turn dialogues. 
E4 pointed out, \qt{each change feels like starting over. I need a way to go back to tweak some [elements], but I don't want to touch certain ones I already feel good about.}
As E2 explained, \qt{I have to copy and paste descriptions into every prompt, just to keep that part, but even so, it often get changed.} 
Without a way to selectively preserve certain visual information and design decisions, designers are forced to manage these iterations manually, making the design refinements inefficient and the creative flow disrupted.

\subsubsection{Design Goals}
To tackle the identified challenges, we articulate the following key design goals to drive the design of \textsc{\sysname{}}. 

\textbf{DG1: Support externalization of selective focus on primitive elements.}
Designers often focus on specific elements in reference images, but current systems require uploading entire images and verbally explaining their focus (C1). 
This creates a gap between what designers attend to and what the system processes. 
To bridge this gap, designers should be allowed to externalize their selective focus of elements into visual representations.
This might involve allowing easy annotation, grouping, and flexible organization of these elements.
By making these elements tangible, we aim to enable designers to interact with them directly, facilitating both their cognitive process and communication with AI at the element level.

\textbf{DG2: Enable spatial management and visual communication of element relationships.}
Designers view elements as interdependent in a design, but articulating relationships such as scale, hierarchy, and spatial proximity is challenging (C2). 
As such, the user interaction should provide a flexible 2D workspace where designers can visually arrange and manipulate elements, defining relationships intuitively, and reducing reliance on verbal descriptions.
The goal is to establish a shared visual structure that allows designers to clearly define the composition while enabling AI to accurately interpret and understand it, improving communication of complex elemental compositions.

\textbf{DG3:  Facilitate element reuse and iterative refinement.}
Designers struggle with the linear nature of current conversational text-to-image tools, which lack mechanisms for selectively preserving or refining elements across iterations (C3). 
To address this, designers should be able to easily reuse individual elements and partial configurations from previous versions to reduce repetitive manual work.
With such a reusing mechanism, designers could explore different design variations more efficiently.


\subsection{Early Prototyping: Co-designing with a Designer}
\label{sec:design_s2}
\subsubsection{Procedure}
In the early stages of our project, we engaged in a four-month co-design process with an expert designer, who has over eight years of graphic design experience. 
We held weekly 30-minute design meetings. 
During this phase, we collaboratively created early low-fidelity mock-ups using sketches and iteratively built non-functional prototypes in Figma. 
This collaboration focused on defining core components of \textsc{\sysname{}} and basic features in \sysname{}.
Given her commitment, we include her as a coauthor. 

\subsubsection{Design Outcomes}
We defined two key aspects of the \textsc{\sysname{}} interaction paradigm: 1) reifying \rev{\cite{beaudouin2000reification}} design elements into \emph{design tokens}; and 2) enabling direct manipulation \rev{\cite{shneiderman1983direct}} on tokens, constructing the \emph{visual lexicon}, to specify relationships. 
We identified two types of design tokens: \textit{visual} and \textit{textual}. Among the visual tokens, we included three core elements: subject, style, and color. 
We also defined five essential manipulation capabilities: \textit{drag-and-drop}, \textit{move}, \textit{resize}, \textit{group}, and \textit{link} tokens. 
The interface operationalizing this paradigm was structured into three panels: 1) a mood board panel for organizing reference images and creating tokens, 2) a token manipulation panel for building relationships, and 3) a history panel to track versions. 
These design decisions led to the development of an initial working prototype.
Details will be described in Section~\ref{sec:paradigm}.

\subsection{Prototype Iteration: Involving Another Six Designers}
\label{sec:design_s3}
\subsubsection{Participants and Procedure.}
To refine the initial design of \textsc{\sysname{}} and the early prototype, we recruited six additional designers with over one year of graphic design experience, each having used at least one text-to-image generation tool more than five times in the past three months. 
We began by walking them through the initial prototype and explaining how to interact with the system.
Using our prepared reference images, we asked them to explore the system and generate multiple designs. 
Designers used a think-aloud method to inform us when they encountered difficulties or desired alternative functionality. 
At the end of the session, participants provided feedback and discussed their overall experience and suggestions for improvement. 
The entire study lasted about 45 minutes, and participants were compensated with the equivalent of \$20 CAD.

\subsubsection{Feedback and Design Refinements.}
Designers identified several inadequacies in the current \textsc{\sysname{}} design and suggested immediate feedback.
Based on their suggestions, we strengthened the visual association between design tokens and their original imagery to improve clarity. 
Additionally, we introduced a \textit{cross-referencing} feature to allow for more effective descriptions of relationships between subject tokens. 
Designers also expressed the need to accommodate both concrete instructions and abstract imagination, so we added an \emph{imaginative token} to the interaction vocabulary. 
These refinements helped make the \textsc{\sysname{}} more expressive.
Details will be described in Section~\ref{sec:paradigm}.

%% file: texfiles/4-duplo.tex
\section{\sysname{}: A Visual-Centric Interaction Paradigm}
\label{sec:paradigm}


In this section, we explain the design decisions and rationale behind \textsc{\sysname{}}, a \emph{visual-centric} interaction paradigm that enables users to express design intent through \emph{direct manipulation on design tokens}.

\subsection{Design Tokens: Specifying What Elements to Use}
We introduce \emph{design tokens} as the externalizations of designers' attended design elements (DG1), reifying~\cite{beaudouin2000reification} the abstract visual information into concrete first-class graphical objects that can be directly manipulated and reused.



\subsubsection{Token types: Being polymorphic to ensure expressiveness and extensibility.}
\begin{figure}[t]
    \includegraphics[width=\linewidth]{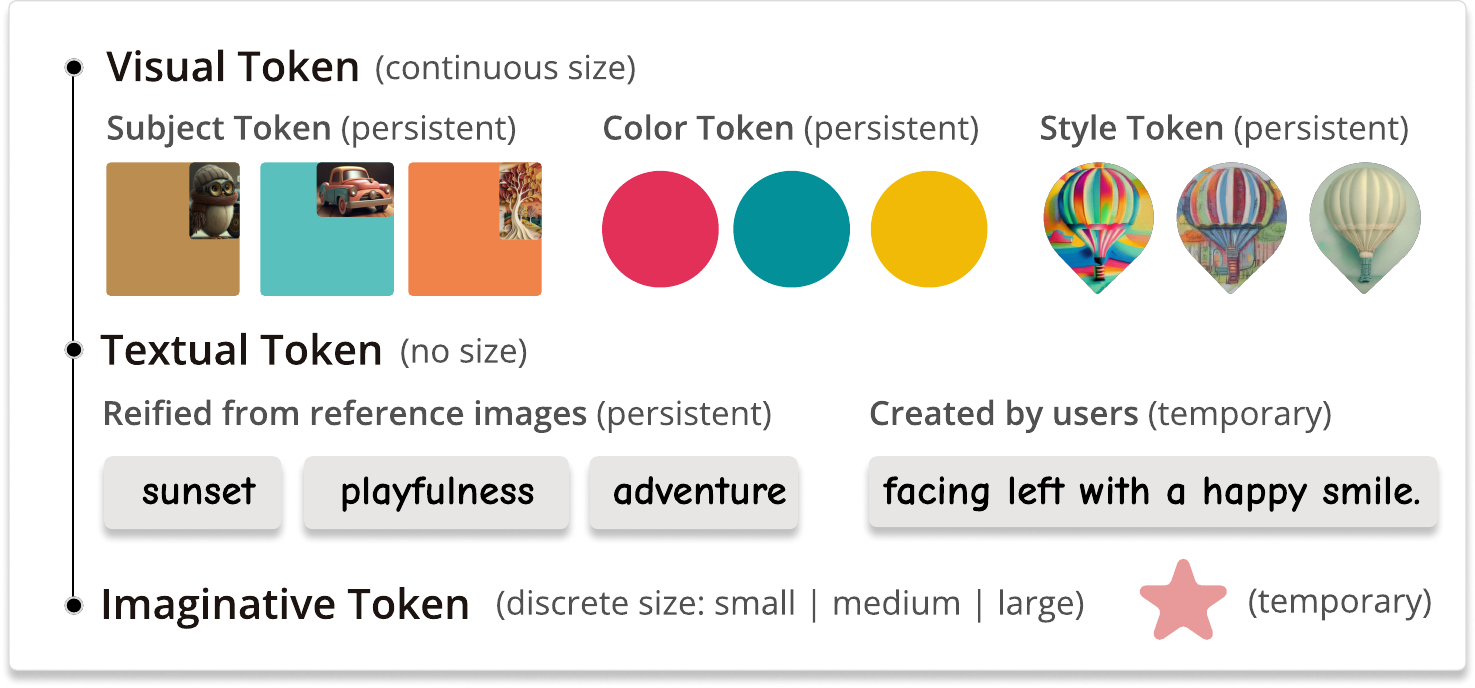}
    \vspace{-6mm}
    \caption{The definition of design tokens in \textsc{\sysname{}}: visual, textual, and imaginative tokens. Each type of tokens has their own appearances and life-cycles.}
    \Description{The figure defines design tokens in Brickify, categorizing them into visual, textual, and imaginative tokens, each with distinct appearances and life cycles. Visual tokens have continuous sizes and are persistent, including subject tokens represented by images, color tokens depicted as solid circles in different colors, and style tokens illustrated by balloon-like images showcasing different artistic styles. Textual tokens have no size and are divided into two types: those reified from reference images, which are persistent (e.g., ``sunset,'' ``playfulness,'' ``adventure''), and those created by users, which are temporary (e.g., ``facing left with a happy smile''). Imaginative tokens have discrete sizes (small, medium, large) and are temporary, exemplified by a pink star shape. The figure highlights how these tokens function within Brickify, each serving a different role in the design process.}
    \label{fig:tokens}
\end{figure}

A key design insight in \textsc{\sysname{}} is that \emph{all types of design elements and intentions should be regarded as tokens}.
The goal is not to support a complete set of all possible design elements but to build an \emph{extensible} paradigm with the affordance to accommodate different types. 
Such \emph{polymorphism}~\cite{beaudouin2000reification} is essential for maintaining a simple interface with consistent interaction logic.
During the early prototyping (S2), the designer expressed the desire for precise control over style, colors, and subject identity.
Later in the prototype iteration stage (S3), participants added that they also appreciated the model's hallucinations for certain details. 
For instance, one participant noted, \qt{I will leave it to the model to decide how an exact \textit{`joyful'} facial expression looks like.} 
Thus, we categorize design tokens into three types: \emph{visual, textual,} and \emph{imaginative} (Fig.~\ref{fig:tokens}).

\vspace{1mm}

\begin{itemize}
    \item[] \textbf{\textsc{Visual} token} carries the visual information such as the \textsc{subject}, \textsc{style}, and \textsc{color}, reified from reference images.
    \item[] \textbf{\textsc{Textual} token} complements visual tokens by conveying information that is easier to express through language, such as adjectives for emotions or verbs for gestures.
    \item[] \textbf{\textsc{Imaginative} token} mediates the initiative between designers and models, indicating \emph{where} the model should intervene and \emph{how much} imagination is needed.
\end{itemize}

\vspace{1mm}

\subsubsection{Token appearances: Balancing fidelity with re-envisioning potential.}

The design tokens can be regarded as a visual abstraction depicting the elements graphically.
When designing their appearance, we balanced between \emph{fidelity} and \emph{re-envisioning potential}. 
Tokens need to be visually distinct, allowing users to easily identify what element they represent while retaining enough abstraction for designers to re-imagine them in new contexts.

In early co-design sessions (S2), we used geometric shapes to represent different elements, \eg, rectangles for subjects, circles for colors, and filled rectangles with different colors to distinguish between subjects. 
Hovering over a token would highlight the original source in the reference image. 
However, in the later prototype iteration stage (S3), we found that as the number of subjects grew and design complexity increased, participants struggled to track which token represented which subject, frequently switching between the mood board and token manipulation panels to confirm identities.
To address this, we refined the subject tokens by attaching a small cropped image of the subject to the token's corner. 
For style tokens, we represented them by transferring the style to a standard image.

\subsubsection{Token life-cycles: Offering both persistent ones for reuse and temporary ones to avoid overwhelm.}
Towards facilitating element reuse and iterative refinement (DG3), we make a distinction between \emph{persistent} and \emph{temporary} tokens. 
Persistent tokens are used for core content elements that are repeatedly referenced throughout the design process, while temporary tokens represent contextual details or single-use modifications.

Designers typically plan a design by considering content, construction, and context. 
For instance, \qt{two girls \emph{(content)} next to each other in the center of the image (\emph{construction)} are dancing surrounded by flowers \emph{(context)}}.
Designers often explore alternative design possibilities by altering the construction or context while keeping the same visual elements. 
To support this process, content-related tokens are persistent for \emph{reuse} across different design variations, while context-related tokens remain temporary to avoid clutter. 
We implemented this distinction by separating token creation into two panels: tokens created in the mood board panel are persistent, with each use being a copy of the original, ensuring the flexibility of reuse. 
In contrast, tokens in the manipulation panel are temporary and deleted when the panel is cleared to prevent interface clutter.

\subsection{Direct Manipulation on Tokens: Expressing How to Construct Elements}
Direct manipulation~\cite{shneiderman1983direct} has long been integral to designers' workflows, especially for rapid prototyping and visual planning. 
In \textsc{\sysname{}}, users express \emph{how} they want to construct elements through direct manipulations on design tokens (DG2).

\begin{figure*}[tbp]
    \centering
    \includegraphics[width=\linewidth]{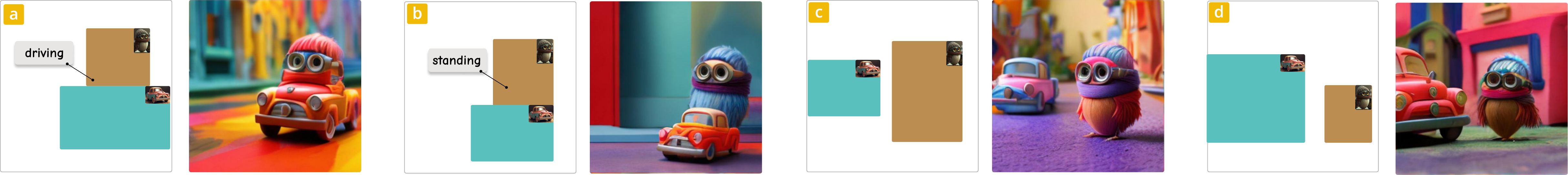}
    \vspace{-7mm}
    \caption{Demonstration of exploring different compositions through direct manipulation on design tokens. (a)–(d) show how adjusting sizes and positions of the owl and car tokens changes their relationships in the outcomes.}
    \label{fig:direct_manipulation}
    \Description{The figure demonstrates how different compositions can be explored through direct manipulation of design tokens, specifically adjusting the sizes and positions of an owl and a car token to alter their relationships in the generated outcomes. On the left, subfigure (a) shows a token representation where the owl is positioned above the car with the label ``driving''. The corresponding generated image depicts an owl sitting inside a small cartoonish red car, suggesting a driving position. Next, subfigure (b) presents a different arrangement where the owl is positioned separately from the car with the label ``standing''. The generated image reflects this change, showing the owl standing on the ground next to the car rather than inside it. In subfigure (c), the owl token is significantly larger than the car token, suggesting a shift in perspective. The generated image portrays the owl as a dominant figure, standing near the car in an exaggerated scale. Finally, subfigure (d) positions the car and owl tokens farther apart, leading to an output where the owl is standing independently on the street while the car remains in the background.
    }
\end{figure*}

\subsubsection{Intuitive actions to reflect intentions.}
The mapping between intention and action should be intuitive, allowing users to engage through \emph{technical reasoning}~\cite{osiurak2009unusual} rather than \emph{procedural learning}.
In collaboration with designers, we defined the following actions to reflect design intent, constructing the \emph{visual lexicon} by manipulating on tokens, as shown in Fig.~\ref{fig:direct_manipulation}. 

\vspace{1mm}
\begin{itemize}

\item[] \textbf{\textsc{Drag-and-drop.}}
Users create persistent tokens (subject, style, color, concept) from reference images in the mood board panel. 
To use these tokens, they can drag their copies and drop them onto the token manipulation panel, where each token can be reused multiple times without limits.

\item[] \textbf{\textsc{Move.}}
Users can freely move subject tokens to define the spatial relationships between subjects. 
While other tokens (\eg, color, style, textual) can also be moved, the movement of these tokens is purely for organizational purposes without encoding spatial relationships, as their function is to modify or describe attributes of the subject.

\item[] \textbf{\textsc{Resize.}}
Resizing tokens adjust their scale. 
Color tokens indicate proportional weights (\eg, primary \vs~ secondary colors), while style tokens work similarly. 
Resizing subject tokens specifies their sizes in the output, and resizing imaginative tokens controls the extent of AI-imagined details. Textual tokens cannot be resized.

\item[] \textbf{\textsc{Group.}}
Grouping tokens helps manage multiple elements easily. 
For example, users can group 3-5 colors into a color theme or apply several colors to a single subject. 

\item[] \textbf{\textsc{Link.}}
Design elements are often interconnected. 
For example, a color token can be linked to a specific subject, applying only to that subject, or left unlinked to apply globally. 
Links specify the relationships between tokens, such as binding colors or textual descriptions to subjects.

\item[] \textbf{\textsc{Cross-reference.}}
Subject tokens often reference one another to specify certain relationships. 
For example, in the phrase \qt{an owl is driving the car}, describing the owl's behavior using a textual token requires referencing the car's token.
To cross-reference, users can assign a name to a token to refer and tag the name in a textual token.
\end{itemize}


\subsubsection{Flexible action reuse.}
In addition to reusing tokens, designers should be able to reuse their previous actions to explore alternative design paths (DG3). 
Since actions are reflected in the construction of design tokens, the visual lexicon, we support action reuse by recording each lexicon created. 
Designers can refine their work based on this visual lexicon rather than redoing previous actions, enabling more efficient iteration and exploration.

\subsubsection{Intermediate action outcomes.}
During prototype refinement (S3), designers expressed the need for immediate feedback on their actions to assess how design tokens respond and evaluate the results. 
However, the current generative models have noticeable inference times and high computational costs, making instant feedback for every action impractical. 
To address this, we introduced feedback at a higher granularity.
Since the execution process follows a sequence --- first composing the layout, then aligning the style, and finally applying colors --- we provide intermediate results at each step.
As model inference times improve, we envision the possibility of real-time feedback for more responsive interaction.

%% file: texfiles/5-usage-scenario.tex
\section{Usage Scenario}

\begin{figure*}[htbp]
    \centering
    \includegraphics[width=.95\linewidth]{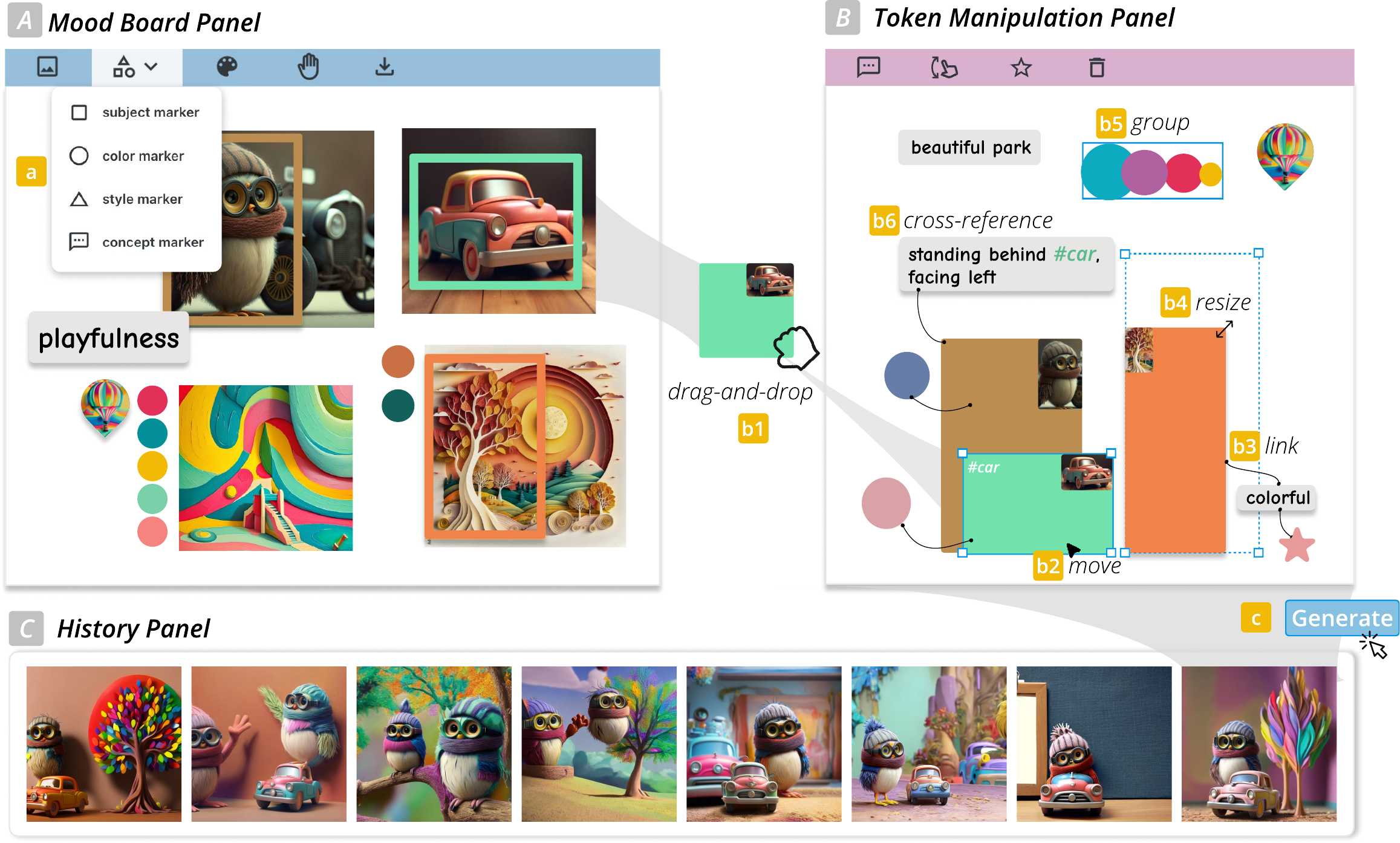}
    \vspace{-3mm}
    \caption{User interface of \sysname{}, consisting of three panels: (A) \emph{Mood Board Panel} for arranging reference images and creating persistent design tokens (subject, color, style, concept), which can be drag-and-dropped (b1) into (B) \emph{Token Manipulation Panel} for direct manipulation (b2 – b6). Clicking the \emph{Generate} button (c), generated results are organized in (C) \emph{History Panel}.}
    \label{fig:interface}
    \Description{The Mood Board Panel (A) on the left serves as a workspace where designers collect reference images, color palettes, and style inspirations. A drop-down menu (labeled a) allows users to mark elements as subject markers, color markers, style markers, or concept markers. Designers can tag reference images with descriptive labels such as ``playfulness'' and annotate specific areas of images, as seen with the car in a green bounding box. Elements can be dragged and dropped (b1) from this panel into the Token Manipulation Panel (B). The Token Manipulation Panel (B) on the right provides tools for structuring design tokens. Designers can group tokens (b5), define cross-references (b6), link elements (b3), move tokens (b2), and resize them (b4). The panel includes tokens extracted from the mood board, such as an owl, a car, and color swatches, which are arranged spatially with annotations to define their relationships. Text descriptions like ``standing behind #car, facing left'' further specify how elements should interact. At the bottom, the History Panel (C) displays previously generated images, allowing designers to review iterations and track changes over time. The Generate button (c) enables users to synthesize new compositions based on their refined token relationships.}
\end{figure*}

Before diving into the design process and the implementation in detail, we walk readers through an example usage scenario to express design intent through \textsc{\sysname{}}.
Imagine Stella, a designer, working on a children's storybook about the adventures of an owl. 
The story follows the owl as he travels in different landscapes with his trusty car and friend.

\textit{Creating design tokens.} 
Stella begins by gathering inspirational images to define the look of the characters and the feel of the scenes and importing them to the Mood Board Panel (Fig.~\ref{fig:interface}A). 
She draws bounding boxes around the owl, car, and tree to create their subject tokens. 
She then creates color and style tokens with the corresponding tools shown in (Fig.~\ref{fig:interface}a). 
To set the thematic tone, she adds a textual token for ``playfulness''.

\textit{Manipulating design tokens.}
She drags and drops (Fig.~\ref{fig:interface}(b1)) the created tokens to the Token Manipulation Panel (Fig.~\ref{fig:interface}B) to build her story (Fig.~\ref{fig:interface}(b2–b6)). 
She imagines the owl parking under a tree, waiting for his friend.
She resizes and positions the owl, car, and tree tokens to define their spatial relationships, then links textual tokens to specify the owl standing behind the car and facing left. 
For the tree, which she imagines as ``colorful'' but undefined, she links an imaginative token to let the AI decide it. 
Stella then groups the color tokens for a cohesive theme and adds a style token. 
For the background, she creates a textual token of ``beautiful park''.

\textit{Reusing design tokens.}
In the next scene, where the owl's friend joins, Stella reuses parts of the previous visual lexicon, making slight adjustments and dragging and dropping another subject token of owl from the Mood Board Panel as his friend. 
By reusing design tokens, she streamlines her workflow and avoids redundant work. 
All generated results and their visual lexicons are organized in the History Panel (Fig.~\ref{fig:interface}C).

%% file: texfiles/6-implementation.tex

\section{\sysname{} System Implementation}

In this section, we explain how \sysname{} extracts primitive design elements from reference images to create design tokens (Section~\ref{sec:token_creation}) and transforms the tokens together with users' actions on tokens into control signals for models to process (Section~\ref{sec:executaion_pipeline}).

\subsection{Design Token Creation}
\label{sec:token_creation}
\subsubsection{Subject token}
Users create a subject token by drawing a bounding box around the desired subject using the subject tool. 
To ensure that generative models accurately capture the visual details specified by users, we employ the \texttt{SAM}~\cite{kirillov2023segment} model to extract segmentation maps and then use the \texttt{Break-A-Scene}~\cite{avrahami2023break} approach to fine-tune the \texttt{Stable Diffusion (v2.1)} model. 
This process learns the subject's visual identity and binds each subject to a specific token within the model for later use.
To accommodate multiple reference images, we concatenate them into one image before fine-tuning because \texttt{Break-A-Scene} can only learn subjects within one single image.

\subsubsection{Color token}
Users can extract color tokens both automatically and manually. 
Using the color tool and clicking on an image automatically extracts five dominant colors as color tokens based on \texttt{K-Means} clustering. 
If the extracted colors do not meet the user's needs, they can click on them to manually change their colors with the color picker. 
If they want to create more color tokens based on one image, they can click again on the image to create a circle with a random color, allowing users to manually select a color with an eyedropper tool to create a color token.

\subsubsection{Style token}
Users indicate their desire to use an image's style by clicking on it with the style tool. 
We leverage \texttt{Style-Align}~\cite{hertz2023style} to transfer the image's style to a standard balloon image, which is then cropped to the marker shape, creating a style token.

\subsubsection{Concept token}
Concept tokens capture the high-level spirit or emotional feeling of an image in textual format. 
When users click on an image with the concept tool, \texttt{GPT-4o} describes the feeling and atmosphere of an image, summarizing it into five keywords.

\begin{figure*}[tbp]
    \centering
    \includegraphics[width=\linewidth]{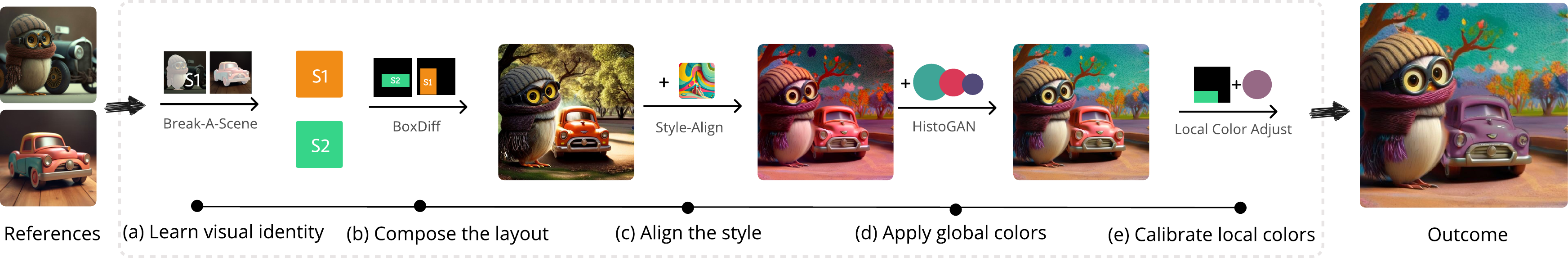}
    \vspace{-7mm}
    \caption{The technical pipeline of \sysname{} interprets and executes the visual lexicon step-by-step, using off-the-shelf methods.}
    \label{fig:pipeline}
    \Description{The pipeline transforms reference images into a final synthesized outcome through a structured process. On the left, the input references include an image of an owl and a car. The first step, Break-A-Scene, analyzes the reference images to learn their visual identity (Step a). Next, BoxDiff is used to compose the layout (Step b) by arranging the extracted elements into a structured composition. The third step, Style-Align, applies a selected style reference to align the style (Step c) with the intended aesthetic. This is followed by HistoGAN, which applies global colors (Step d) by integrating a defined color palette into the generated image. Finally, Local Color Adjust fine-tunes color consistency at a localized level to calibrate local colors (Step e), ensuring that all elements blend seamlessly. The result is a cohesive final outcome on the right, featuring an owl and a car in a visually harmonious scene.}
\end{figure*}

\subsection{Visual Lexicon Execution}
\label{sec:executaion_pipeline}
To ensure a smooth iterative design experience, we selected the approaches for the visual lexicon execution that do not require training or fine-tuning on diffusion models. 
It should be noted that the field of computer vision evolves rapidly and methods could be replaced as better solutions emerge, our goal is to provide a feasible technical pipeline for executing the visual lexicon users create.
The execution consists of four primary steps: handling layout, style, global colors, and local colors (Fig.\ref{fig:pipeline}).
This order is deliberately designed to prevent visual effects from being overridden. 
For example, handling the style inevitably changes the colors to some degree, so color adjustments must come afterward. Similarly, global colors are handled before local colors.

\subsubsection{Extend keywords description}
Users often use only keywords in textual tokens, but diffusion models perform better with complete sentences as prompts. 
We thus extend textual tokens into sentences using \texttt{GPT-4o}. 
The size of the imagination token determines the level of detail added, with three levels: small, medium, and large. 
If no imagination tokens are used, only factual information is stated without any added imagination.

\subsubsection{Compose the layout}
To compose subjects into the desired layout, we use \texttt{BoxDiff}~\cite{xie2023boxdiff}, which constrains image generation with spatial control guidance. 
For the foreground, it takes the subject token placements as bounding boxes and related textual keywords describing each bounding box as input, triggering the special tokens in the pre-trained diffusion model fine-tuned by \texttt{Break-A-Scene} to generate subjects according to the specified layout.
It also handles the background generation with the given text description.

\subsubsection{Align the global style}
By default, the image composed in the layout at the previous step is in a realistic, natural style. 
To align it with the user-specified style token, we use the \texttt{Style-Align}~\cite{hertz2023style} method. 
\texttt{Style-Align} performs shared self-attention with the reference style image at each diffusion denoising step to achieve style alignment during the image reconstruction process.

\subsubsection{Apply global colors}
To apply the global color palette to the image, we employ the \texttt{HistoGan}~\cite{afifi2021histogan} to recolorize images based on the given palette. 
Specifically, we use the pre-trained checkpoint of \texttt{Universal model-0}. 
This method is based on generative adversarial networks (GANs) and optimizes the recolored image to match the color proportion distribution by projecting color histogram features into the model’s latent space. 

\subsubsection{Calibrate local colors}
Lastly, we handle the color tokens attached to local subjects. 
\texttt{Blended Latent Diffusion}~\cite{avrahami2023blended} is leveraged to perform local color modifications on the image. 
This method targets the local editing of generic images, where the desired edits are confined to a user-provided mask without touching the rest.


%% file: texfiles/7-study-design.tex
\section{User Study Design}
\label{sec:study_design}
We conducted an in-lab user study to evaluate the effectiveness of \textsc{\sysname{}} in intent expression and users' experience when interacting with the system of \sysname{}.
The study involved two tasks with the same set of participants.

\subsection{Participants}
We recruited 12 experienced designers via social media and mailing lists. 
All participants hold a formal design degree and have more than 3 years of experience in graphic design. 
Participants rated their frequency of using generative AI tools for text-to-image generation on a 5-point Likert scale (1 = ``never'' to 5 = ``very often''), with an average rating of 3.33.
Their detailed demographic information is listed in Appendix~\ref{tab:evaluation_demographic}.
Participants took part in the study remotely. 
All study sessions were audio and video recorded. 
The entire study lasted about 75 - 100 minutes, and participants were compensated with the equivalent of \$30 CAD.
The study was approved by the university’s ethics review board.

\subsection{Study 1: Interaction Paradigm Comparison}
Study 1 uses a replication task to simulate a scenario where designers have a well-developed idea in mind. 
The goal is to answer the research question (\textbf{RQ1}): How does the \emph{visual-centric} interaction paradigm of \textsc{\sysname{}} compare to the \emph{textual-centric} paradigm in terms of clarity, mental effort, and time investment for expressing design intent? 

\begin{figure}[htbp]
    \centering
    \includegraphics[width=\linewidth]{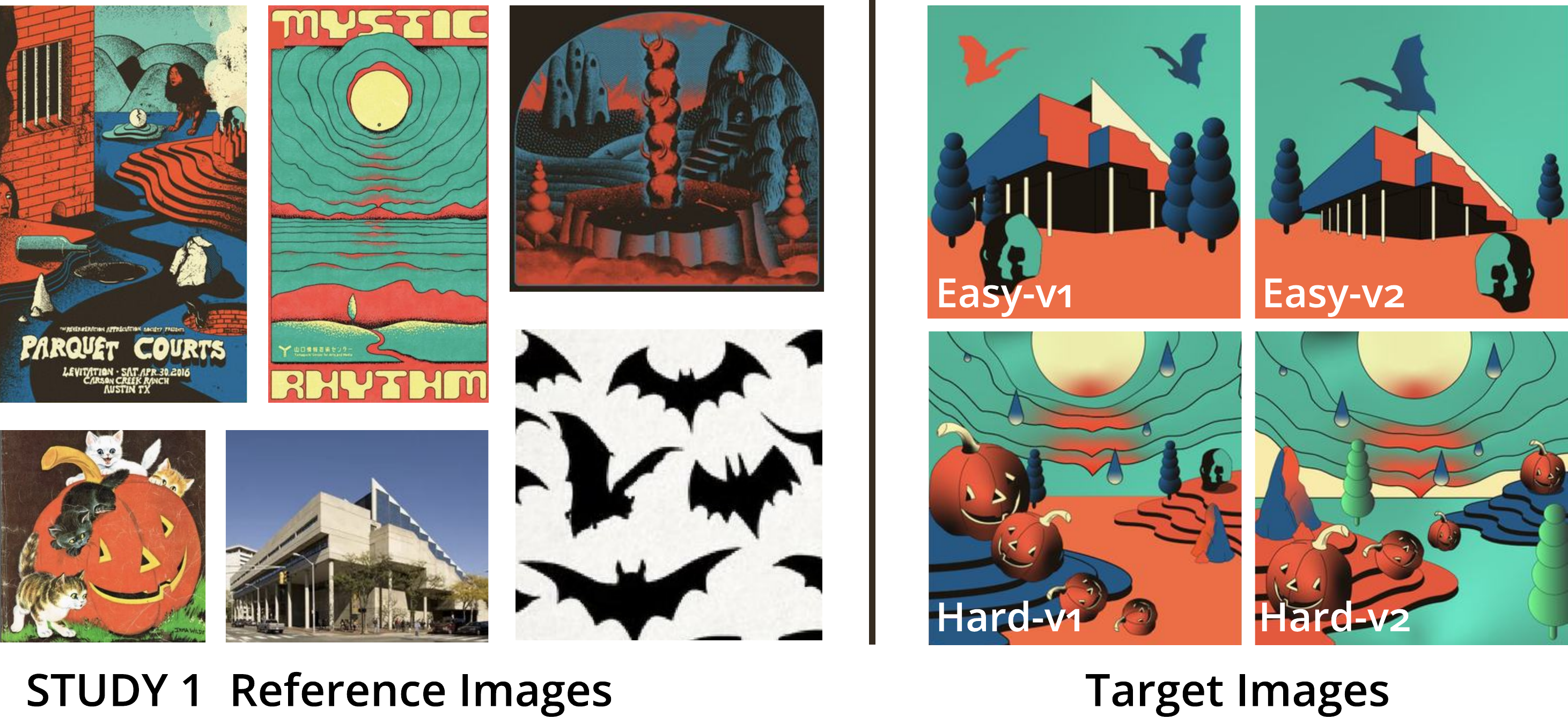}
    \vspace{-6mm}
    \caption{Reference and target images for Study 1. For each condition (\textsc{Easy} and \textsc{Hard}), users work with two versions: one created from scratch and a second tweaked version where users adjust their original description to match the modified target image.}
    \label{fig:study1_images}
    \Description{The figure presents reference and target images used in Study 1 to evaluate user interactions under different conditions. On the left, the reference images consist of a diverse set of visuals, including stylized illustrations, architectural photos, and graphic patterns featuring elements such as landscapes, pumpkins, and bats. On the right, the target images are divided into two conditions: Easy and Hard, each with two versions (v1 and v2). In the Easy condition, the target images depict a simplified scene with geometric structures, trees, and birds. In the Hard condition, the images are more complex, incorporating multiple pumpkins, detailed shading, and richer compositions. Users work with two variations for each condition: one generated from scratch and another requiring them to tweak their original descriptions to align with a modified target image. }
    \vspace{-3mm}
\end{figure}

\subsubsection{Experimental Design}
We use a $2 \times 2$ within-subject design with two primary factors: \textsc{Technique} (\textsc{\sysname{}} or \textsc{Baseline}) $\times$ \textsc{Difficulty} (\textsc{Easy} or \textsc{Hard}).
The \textsc{\sysname{}} condition, as described in Section~\ref{sec:paradigm}, is our proposed interaction paradigm, accessed through the \sysname{} interface shown in Figure~\ref{fig:interface}, with the \textit{Generate} button disabled. 
In the \textsc{Baseline} condition, participants describe their design intentions by typing textual prompts in a Google Doc, with the ability to refer to reference images by their provided names.
We avoided using existing commercial text-to-image interfaces like Dall-E or MidJourney as baselines due to their differing interaction designs --- Dall-E allows uploading an unlimited number of images but lacks a clear mechanism to reference them while MidJourney uses command-style prompts and requires images to be in URL format. 
To ensure fairness for participants familiar with different tools, we thus provided this general textual-centric prompting method reflecting common generative model interactions as a baseline.

The \textsc{Difficulty} levels (\textsc{Easy} and \textsc{Hard}) are determined based on the number of visual elements and the complexity of their compositions. 
To establish these conditions, an expert designer selected elements from a set of reference images and created initial versions for both difficulty levels. 
For the \textsc{Easy} condition, the design involved fewer elements and simpler compositions, while the \textsc{Hard} condition included more elements with more intricate arrangements. 
The designer then further refined the compositions to create a second version for each condition, as shown in Fig.~\ref{fig:study1_images}. 

Participants were asked to express their design intent for the target images across the two \textsc{Difficulty} levels (\textsc{Easy} and \textsc{Hard}) using two interaction paradigms: \textsc{\sysname{}} (visual-centric) and \textsc{Baseline} (textual-centric). 
For each \textsc{Difficulty} level, participants first described the design of a target image (\textsc{Easy-v1} or \textsc{Hard-v1}), then refined their expression to produce a modified version (\textsc{Easy-v2} or \textsc{Hard-v2}), simulating a real-world design refinement process.

\subsubsection{Measurements}



\rev{
To assess whether user expressions sufficiently described the target image, we adopted a human-evaluation approach by recruiting three external raters to assess all participants’ expressions under both \textsc{Techniques}.
Raters self-reported frequent use of text-to-image generation tools ($M = 4.33, SD = 0.58$; scale: 1 = ``never'' to 5 = ``very often'') and demonstrated being knowledgeable in prompt engineering ($M = 3.67, SD = 0.58$; scale: 1 = ``no experience'' to 5 = ``expert'').
The three raters performed the evaluation independently on five 7-point Likert Items (\ie, element coverage, size clarity, position clarity, style clarity, and color clarity) followed the predefined rubric (see Appendix~\ref{sec:rubric}). 
The expressions were order-randomized for each rater. 
The rating process took around 2 hours and the raters received \$50 CAD for their time.

We chose not to use AI models to directly execute expressions and generate final outcomes to compare due to the following reasons. 
First, there is no off-the-shelf techniques built on top of Stable Diffusion 2.1 (the same base model in our technique) that resemble \textsc{Baseline} --- taking multiple references as input and leveraging their subjects and styles --- to make it comparable. 
Second, as in this study we focus on the expressivity aspect, uncertainty and complexity in the process of execution for AI models may introduce compounding factors unrelated to the quality of user expressions.
Therefore, we instead rely on human raters, whose evaluations could more effectively reflect the quality of the expressions from the message receiver perspective.
}

To this end, the measurements for Study 1 included 1) participants' responses to five questions evaluating intent expression, \rev{2) external ratings for participants' expression in different conditions}, 3) task completion times for both the initial and refined design versions,  4) participants' preferences between the two \textsc{Technique}s, and 5) self-reported cognitive load during the tasks.



\subsection{Study 2: \sysname{} Exploration}
Study 2 is an open-ended task, without comparison with other systems, designed to explore the research question (\textbf{RQ2}): How does \textsc{\sysname{}} influence users' creative exploration when they start without a clear intent?

\begin{figure}[htbp]
    \includegraphics[width=\linewidth]{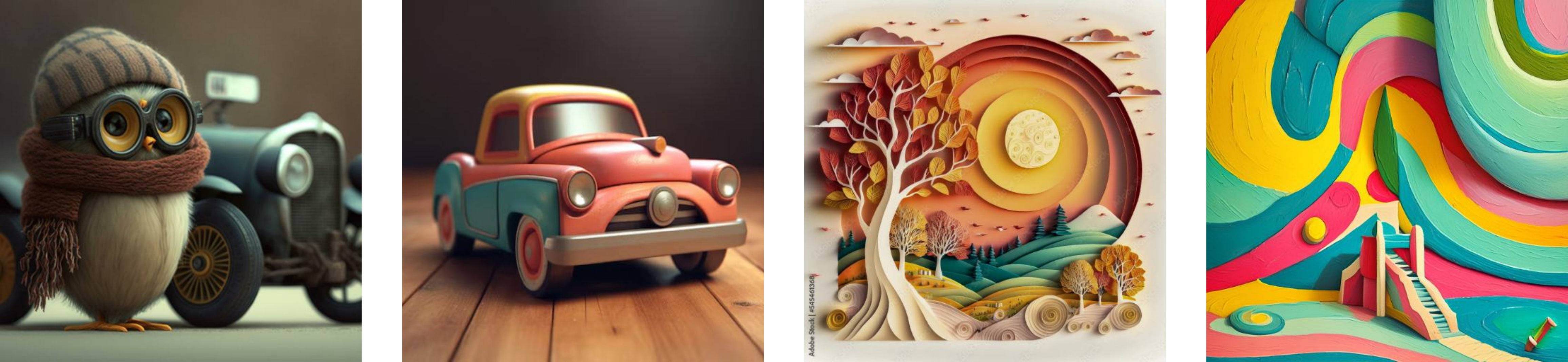}
    \vspace{-5mm}
    \caption{Reference images used in Study 2.}
    \Description{The figure showcases the reference images used in Study 2, consisting of four visually distinct compositions. The first image features an owl wearing a scarf and goggles, standing beside a small car, evoking a cozy and adventurous aesthetic. The second image is a stylized vintage car placed on a wooden surface, with a soft lighting effect enhancing its playful and retro feel. The third image is an artistic paper-cut-style landscape, depicting a tree, mountains, and a large sun with intricate layered textures. The fourth image presents a vibrant abstract scene with bold, curving shapes in bright colors, creating a dynamic and surreal environment. }
    \label{fig:reference_study2}
    \vspace{-2mm}
\end{figure}

\subsubsection{Task design}
In this task, participants assumed the role of junior graphic designers tasked with creating a graphic series for a children's storybook about the adventures of an owl. 
The senior designer provided four reference images (Figure~\ref{fig:reference_study2}) to define the visual characteristics. 
Participants were asked to create three images depicting scenes where the owl, with or without his friend and car, embarks on an adventure. 
The task required maintaining visual consistency across all images. 
There was no time limit, and participants worked until they felt their designs were complete. 

\subsection{Procedure}
After signing the consent form, participants were given an overview of the study procedure, duration, and data collection details. 
The studies were conducted remotely via Zoom, with participants accessing \sysname{} through the web browser. 
A brief training session, including a toy example, was provided to demonstrate the use of \sysname{} and explain the text-based prompting in the \textsc{Baseline} condition. 
Participants could familiarize themselves with the tools before starting the tasks.

Participants began with Study 1, completing both \textsc{Easy} and \textsc{Hard} tasks using \textsc{\sysname{}} and the \textsc{Baseline}, with no time limits. 
The sequence of the four trials (\textsc{Baseline}-\textsc{Easy}, \textsc{Baseline}-\textsc{Hard}, \textsc{\sysname{}}-\textsc{Easy}, \textsc{\sysname{}}-\textsc{Hard}) was randomly assigned across participants, as a full counter-balancing was not feasible. 
After each trial, participants filled out questionnaires to rate their intent expression experience. 
Upon completing Study 1, they filled out the questionnaire on their preference and cognitive load on a 7-point Likert Scale.
Next, participants proceeded to Study 2. 
After completing this task, they rated their experience with \sysname{} using a post-study questionnaire for Creativity Support Index (CSI)~\cite{cherry2014quantifying}. 
\rev{
A semi-structured interview was conducted to gather feedback on participants' experiences with \sysname{}. 
Participants were encouraged to share general comments on any aspect of the study and then prompted on specific aspects, including interface usability frustrations, challenging intention expression cases, prior difficulties with text-centric GenAI tools, and whether similar issues arose with \sysname{}.
They were also asked about the system's impact on their approach to solve design problems, exploration of design options, and suggestions for improvement. 
Observations noted by the experimenter during the session were also discussed.
}

\subsection{Data Analysis}


To analyze the qualitative feedback, we analyzed interviews using thematic analysis, employing both inductive and deductive approaches. 
Two researchers collaboratively analyzed and open-coded the transcribed interviewees’ responses, employing affinity diagramming to sort the initial codes onto cards. 
Then, they discussed and reconciled any discrepancies in the coding process to ensure a consistent and accurate representation of participants’ perspectives. 
Through iterative discussions and the organization of these codes, we identified a number of recurring patterns and themes within the interview data.

%% file: texfiles/8-study-results.tex
\section{User Study Results}
\label{sec:study_result}

\subsection{Self-Rated Design Intent Expression}

\begin{figure*}[htbp]
    \centering
    \begin{subfigure}[]{\textwidth}
    \centering
        \includegraphics[width=.6\textwidth]{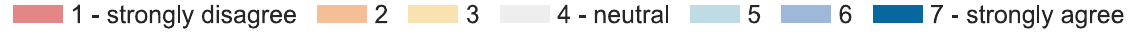}
        \label{fig:intent1}
        \vspace{2mm}
    \end{subfigure}
    \vspace{2mm}
    \begin{subfigure}[]{\textwidth}
        \includegraphics[width=\textwidth]{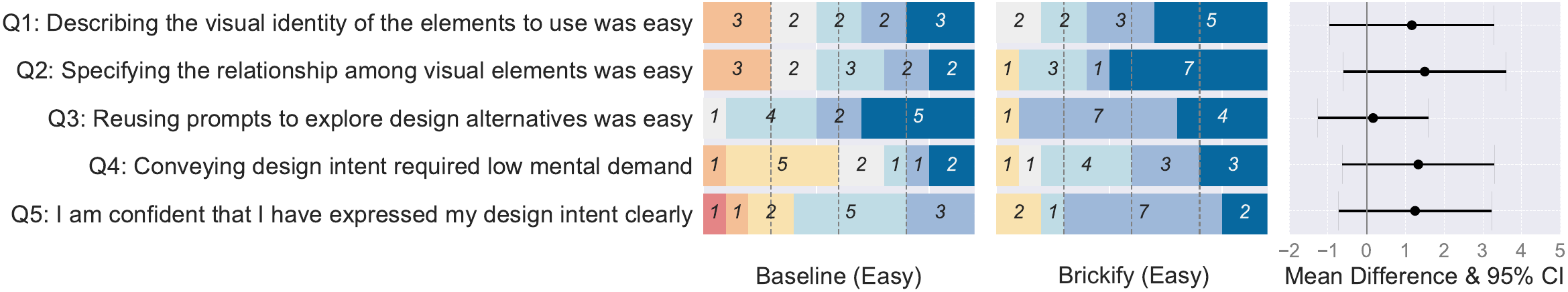}
        \label{fig:intent2}
        \vspace{-3mm}
    \end{subfigure}
    \begin{subfigure}[]{\textwidth}
        \includegraphics[width=\textwidth]{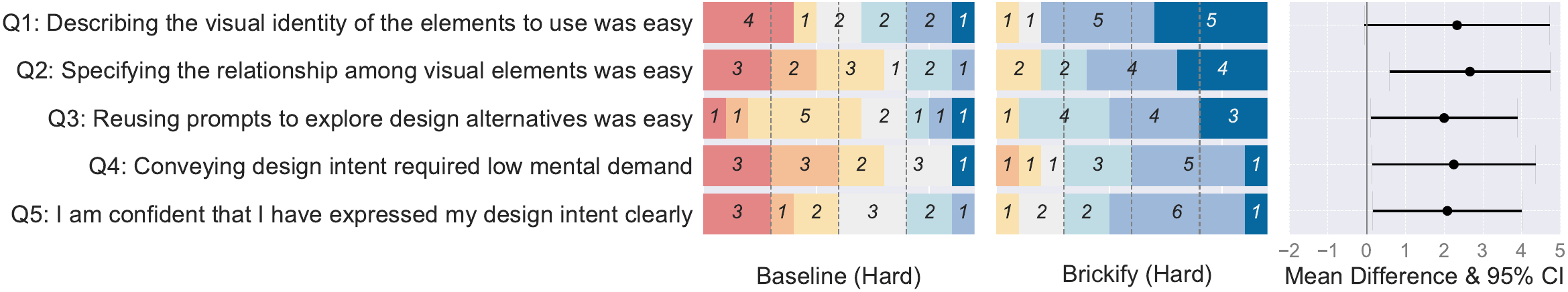}
        \label{fig:intent3}
    \end{subfigure}
    \vspace{-6mm}
    \caption{Participants' response for Study 1 when rating the 7-point statements for \textsc{Baseline} and \textsc{\sysname{}} interaction paradigm under \textsc{Easy} and \textsc{Hard} conditions. Dots are the mean differences of \textsc{\sysname{}} compared to \textsc{Baseline}.}
    \label{fig:intent_results}
    \Description{Participants' response for Study 1 when rating the 7-point statements for \textsc{Baseline} and \textsc{\sysname{}} interaction paradigm under \textsc{Easy} and \textsc{Hard} conditions. Dots are the mean differences of \textsc{\sysname{}} compared to \textsc{Baseline}.}
    \vspace{-2mm}
\end{figure*}

\begin{figure*}[h!]
    \centering
    \includegraphics[width=\linewidth]{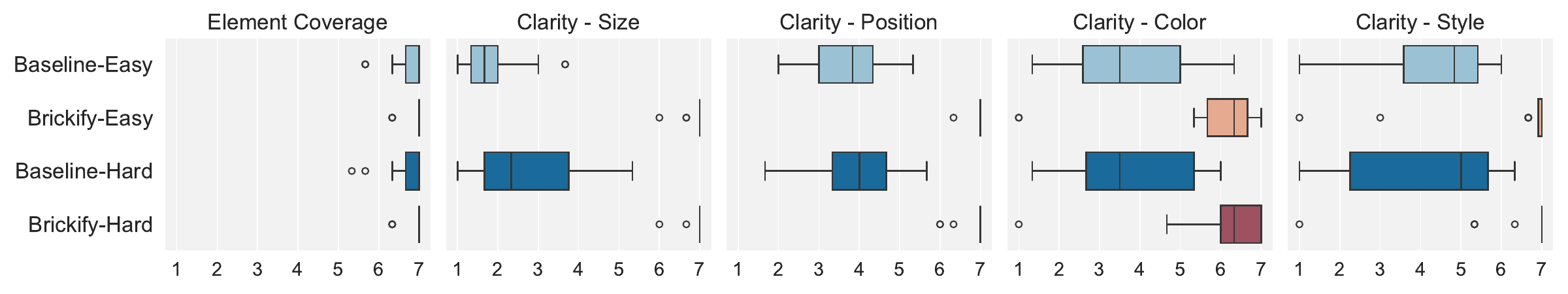}
    \vspace{-7mm}
    \caption{External rating for the quality of expressions that participants produced in \textsc{Baseline} and \textsc{\sysname{}} regarding the element coverage, the clarity of size, position, style, and color.}
    \label{fig:external_rating}
    \Description{External rating for the quality of expressions that participants produced in \textsc{Baseline} and \textsc{\sysname{}} regarding the element coverage, the clarity of size, position, style, and color.}
\end{figure*}

We conducted the non-parametric \textit{Aligned Rank Transform (ART) ANOVA}~\cite{wobbrock2011aligned} statistical analysis on the ordinal Likert-Scale subjective ratings for Study 1 to understand the influence of \textsc{Technique} and \textsc{Difficulty} on users' self-reported design intent expression experience (Q1-Q5).
Results show that there is a significant main effect of \textsc{Technique} on users' self-reported design intent expression experience across all five questions, while \textsc{Difficulty} had significant effects for Q2, Q3, Q4, and Q5 but not for Q1. 
The interaction effect between \textsc{Technique} × \textsc{Difficulty} was significant for Q3 only.
The post-hoc multi-factor contrast tests following the ART-C~\cite{elkin2021aligned} procedure were conducted to identify the exact differences. 
The results show that participants' ratings on Q2, Q4, Q5 were significantly higher for \textsc{\sysname{}} than the Baseline in the \textsc{Easy} task. 
For \textsc{Hard} task, participants' ratings for \textsc{\sysname{}} on all five questions (Q1-Q5) were significantly higher than the Baseline.

The analysis, along with the detailed scores in Fig.~\ref{fig:intent_results}, demonstrates that \textsc{\sysname{}} is more effective than the text-centric interaction paradigm in supporting users' design intent expression, particularly in \emph{higher difficulty} tasks. \textsc{\sysname{}} simplifies describing visual identity, specifying relationships, reusing prompts for alternatives, reduces cognitive load, increases expression confidence, and enhances the overall design expression experience.

\subsection{External-Rated Expression Quality}

\rev{
Figure~\ref{fig:external_rating} shows the external ratings for participants' expressions across five items --- element coverage, clarity of size, position, color, and style (the rubric is shown in Appendix~\ref{sec:rubric}). 
The reliability of the ratings was measured using a two-way random Intraclass Correlation Coefficient (ICC). 
The ICC values for each item ranged from 0.688 to 0.930, with an average of 0.817, indicating acceptable reliability. 
We use the average score from the three raters for each expression for further statistical analysis.
Across different \textsc{Difficulty} levels, participants in both the \textsc{Baseline} and \textsc{\sysname{}} conditions successfully covered most elements in the target images. While \textsc{\sysname{}} received slightly higher scores for element coverage, the difference was not statistically significant.
For the perceived clarity, the two-way ANOVAs indicated that the choice of \textsc{Technique} (\ie, \textsc{Baseline} \vs~ \textsc{\sysname{}}) is the primary factor significantly influencing the clarity of size (\anova{1}{92}{892.15}{.001}), position (\anova{1}{92}{501.02}{.001}), color (\anova{1}{92}{58.14}{.001}), and style (\anova{1}{92}{51.80}{.001}) regardless of task difficulty.
These results suggest \textsc{\sysname{}} provides a more effective approach for reducing the ambiguity in intent expression.
}

\subsection{Task Completion Time}
\label{sec:completion_time}

    


\begin{figure*}[htbp]
    \centering
    \includegraphics[width=\linewidth]{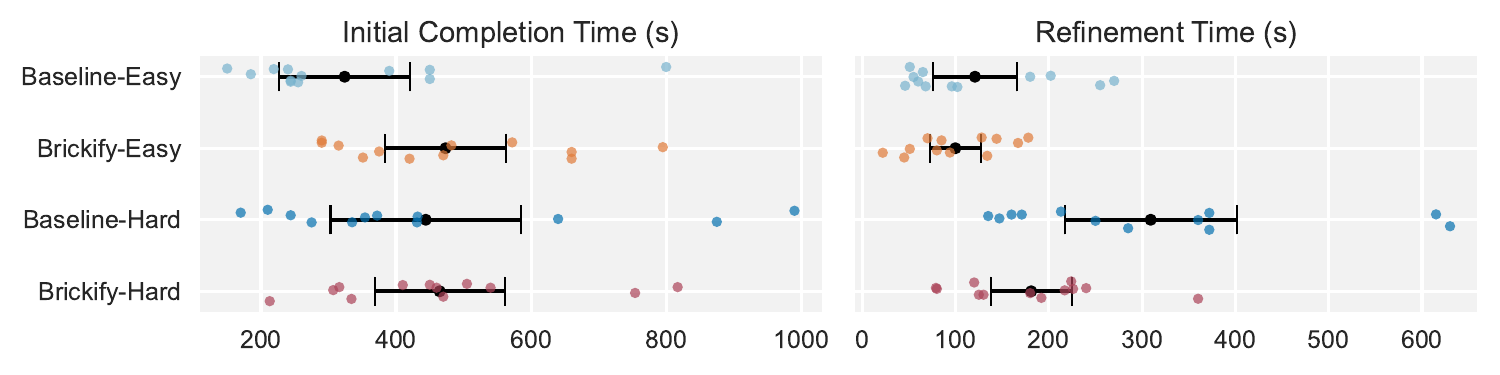}
    \vspace{-6mm}
    \caption{Users' \textsc{Initial Completion Time} (left) and the \textsc{Refinement Time} (right) with \textsc{Baseline} and \textsc{\sysname{}} interaction paradigm under \textsc{Easy} and \textsc{Hard} conditions in Study 1. Black dots are means, bars are 95\%CIs.}
    \label{fig:time}
    \Description{Users' \textsc{Initial Completion Time} (left) and the \textsc{Refinement Time} (right) with \textsc{Baseline} and \textsc{\sysname{}} interaction paradigm under \textsc{Easy} and \textsc{Hard} conditions in Study 1. Black dots are means, bars are 95\%CIs.}
\end{figure*}

\begin{figure*}[htbp]
    \centering
    \begin{subfigure}[]{\textwidth}
        \includegraphics[width=\textwidth]{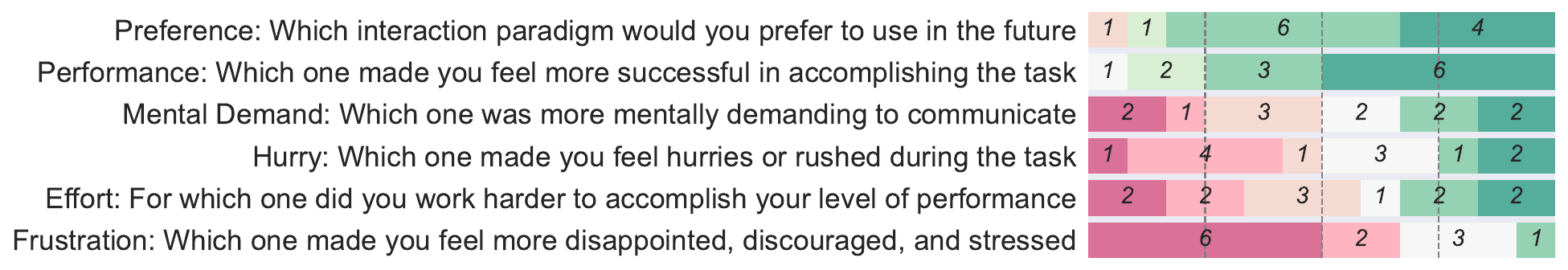}
        \label{fig:nasa1}
        \vspace{-3mm}
    \end{subfigure}

    \begin{subfigure}[]{\textwidth}
    \centering
        \includegraphics[width=.7\textwidth]{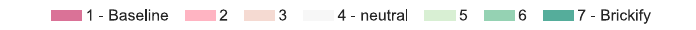}
        \label{fig:nasa2}
        \vspace{2mm}
    \end{subfigure}
    
    \vspace{-5mm}
    \caption{Participants self-reported preference and cognitive load that directly compare \textsc{Baseline} and \textsc{\sysname{}}.}
    \Description{Participants self-reported preference and cognitive load that directly compare \textsc{Baseline} and \textsc{\sysname{}}.}
    \label{fig:nasa}
    \vspace{-2mm}
\end{figure*}

In Study 1, on average, participants took longer to complete the initial version in the \textsc{\sysname{}} condition compared to the \textsc{Baseline} condition ($\overline{t}_{\textsc{Baseline}\_\textsc{Easy}} = 324s$ \vs~ $\overline{t}_{\textsc{\sysname{}}\_\textsc{Easy}} = 474s$, $\overline{t}_{\textsc{Baseline}\_\textsc{Hard}} = 444s$ \vs~ $\overline{t}_{\textsc{\sysname{}}\_\textsc{Hard}} = 465s$), detailed data is shown in Figure~\ref{fig:time}.
However, the differences were not statistically significant.
An ANOVA analysis of the model \textsc{Initial Completion Time} $\sim$ \textsc{Technique} × \textsc{Difficulty} revealed no significant main effects of \textsc{Technique} (\anovans{1}{44}{2.93}) or \textsc{Difficulty} (\anovans{1}{44}{1.44}), and no significant interaction between \textsc{Technique} and \textsc{Difficulty} (\anovans{1}{44}{1.32}).
This result aligns with our expectations, as \textsc{\sysname{}} inherently involves many operations (\eg, creating, dragging, and constructing tokens) that participants have to perform to express their intent from scratch. 
\rev{
The breakdown of \textsc{Initial Completion Time} shows that participants spent approximately one-third ($\overline{t}_{\textsc{Easy}} = 139s$, $\overline{t}_{\textsc{Hard}} = 165s$) of the total time in creating tokens and dragging them to the manipulation panel. 
The remaining time was dedicated to manipulating these tokens to construct the lexicon.
}

When entering the refinement stage, participants, on average, took less time in \textsc{\sysname{}} condition to refine their initial expression to achieve the modified version than \textsc{Baseline} condition ($\overline{t}_{\textsc{Baseline}\_\textsc{Easy}} = 121s$ \vs~ $\overline{t}_{\textsc{\sysname{}}\_\textsc{Easy}} = 99s$, $\overline{t}_{\textsc{Baseline}\_\textsc{Hard}} = 309s$ \vs~ $\overline{t}_{\textsc{\sysname{}}\_\textsc{Hard}} = 181s$).
The ANOVA in the model \textsc{Refinement Time} $\sim$ \textsc{Technique} × \textsc{Difficulty} shows significant main effects of \textsc{Technique} (\anova{1}{44}{5.81}{.05}) and \textsc{Difficulty} (\anova{1}{44}{18.03}{.001}), but no significant \textsc{Technique} × \textsc{Difficulty} interaction (\anovans{1}{44}{2.97}) on \textsc{Refinement Time}.
Post-hoc Tukey HSD tests further show that users spent significantly less \textsc{Refinement Time} in \textsc{\sysname{}} on average by 128 seconds ({$p = .027$}) in \textsc{Hard} condition, while no significant difference found for \textsc{Easy} condition. 
This result can be attributed to the fact that participants did not create new tokens in \textsc{\sysname{}} during the refinement stage; instead, they focused solely on manipulating existing tokens. 
The faster \textsc{Refinement Time} highlights \textsc{\sysname{}}'s strength in enabling quicker and more efficient modifications once the initial design intent is established.


\subsection{User Preference and Cognitive Load}

After completing Study 1, participants were asked to rate their preference between \textsc{\sysname{}} and the Baseline condition, as well as their cognitive load for each condition (Fig.~\ref{fig:nasa}).
Participants (11/12) showed a clear preference for \textsc{\sysname{}} over the Baseline. 
On a 7-point Likert scale (1 = strongly prefer Baseline, 7 = strongly prefer \textsc{\sysname{}}), the mean preference rating of 6.0 was significantly above the neutral midpoint of 4. 
This indicates a strong preference for \textsc{\sysname{}} when participants had a clear design intent to express.

Regarding cognitive load, most participants (11/12) felt they were more successful using \textsc{\sysname{}}. 
Additionally, 7 out of 12 participants reported making less effort, and 8 out of 12 felt less frustrated, suggesting that \textsc{\sysname{}} simplifies the design intention expression and reduces users' frustration.
However, only half of the participants felt reduced mental demand and hurry compared to the Baseline. 
As users engage more deeply in articulating their design intent, they may invest more time and mental effort in decisions such as token placements.
Overall, \textsc{\sysname{}} enhances the design intent expression clarity and reduces frustration but still requires a certain level of mental effort and time commitment to fully engage with design token manipulations.

\subsection{User Behavior}


\begin{figure}[htbp]
    \centering
    \includegraphics[width=.9\linewidth]{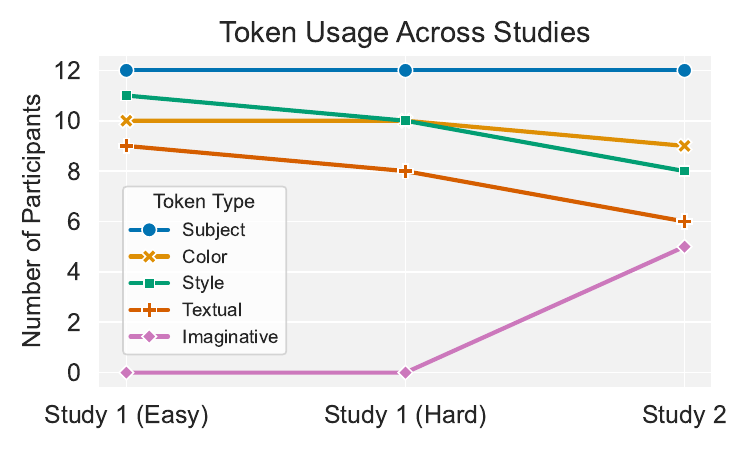}
    \vspace{-3mm}
    \caption{Token usage in \sysname{} across studies, indicating the number of participants have used each token type.}
    \label{fig:token_usage}
    \Description{Token usage in \sysname{} across studies, indicating the number of participants have used each token type.}
    \vspace{-2mm}
\end{figure}

\rev{
\subsubsection{Token Usage}
Figure~\ref{fig:token_usage} illustrates the number of participants using each token type across studies. 
Subject tokens were consistently adopted by all participants across all conditions, while color, style, and textual tokens were used by most participants (more than 8). There was a slightly reduced usage (but still more than half of the participants) in Study 2 for these tokens, because they focused more on shaping the narrative by manipulating subject tokens. 
Notably, no participants used imaginative tokens in Study 1, which may be because of its well-defined scenario with clear targets relied entirely on reference images. 
During Study 2, where the design scenario was more exploratory, five participants 
employed imaginative tokens for (1) enriching background descriptions (P2, P4, P10), (2) modifying subject tokens to adjust visual identity and/or actions, \eg, differentiating the owl’s friend from the owl (P11), and (3) expanding text descriptions for global styles (P4, P9).

\subsubsection{User Strategy}
Across both studies, we observed several strategies shared across participants when interacting with \sysname{} to construct the visual lexicon.
All participants adopted an on-demand approach for token creation, using tokens as needed rather than creating all tokens upfront.
This led to frequent navigation between the mood board panel and the token manipulation panel. 
When constructing the visual lexicon, 9 out of 12 participants prioritized creating and positioning subjects first, followed by adding local colors (if any), and then applying global styles and/or colors. 
The remaining participants began by considering global styles and colors before adding subjects and their local colors. 
We also observed that most participants (N = 10) started with the background or underlying layers and worked progressively toward the foreground. 
As P2 explained in the interview, this approach likely stems from their unconscious habit of working with layer-based logic in tools like Adobe Photoshop and Illustrator: \qt{I always build layer by layer, back to front in Illustrator and it feels natural to follow that order.}
}


\subsection{Self-reported Creativity Support Index}

     

\begin{figure}[htbp]
    \centering
    \includegraphics[width=\linewidth]{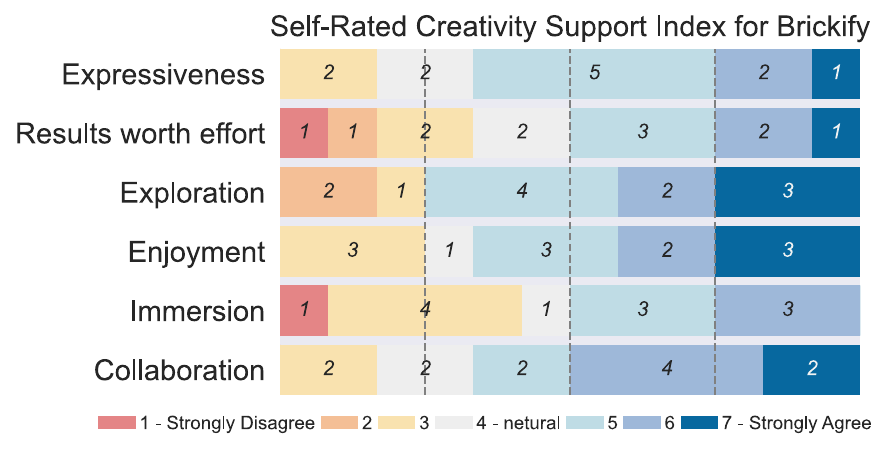}
    \vspace{-5mm}
    \caption{Self-reported Creativity Support Index for \sysname{} after Study 2.}
    \label{fig:csi}
    \Description{Self-reported Creativity Support Index for \sysname{} after Study 2.}
    \vspace{-2mm}
\end{figure}

We utilize the Creativity Support Index (CSI) to measure the degree of creativity support for \sysname{} in the Study 2. 
Since Study 2 does not include a baseline for comparison, we present this self-reported rating as a reference point to better understand users' experiences, rather than drawing definitive conclusions.
Participants rated six creativity support factors on a scale from 1 (strongly disagree) to 7 (strongly agree) shown in Fig.~\ref{fig:csi}: expressiveness, results-worth-effort, exploration, enjoyment, immersion, and collaboration.
Overall, \textsc{\sysname{}} shows strong support for creativity, effectively supports idea exploration, and is generally enjoyable to use.
Although most participants rated positively, one participant strongly disagreed with the immersion factor, which may reflect interface limitations, as noted in an interview where the user suggested frequently used tools to create visual tokens should be directly available in the toolbar rather than in a list.

\begin{figure*}[tbp]
    \centering
    \includegraphics[width=\linewidth]{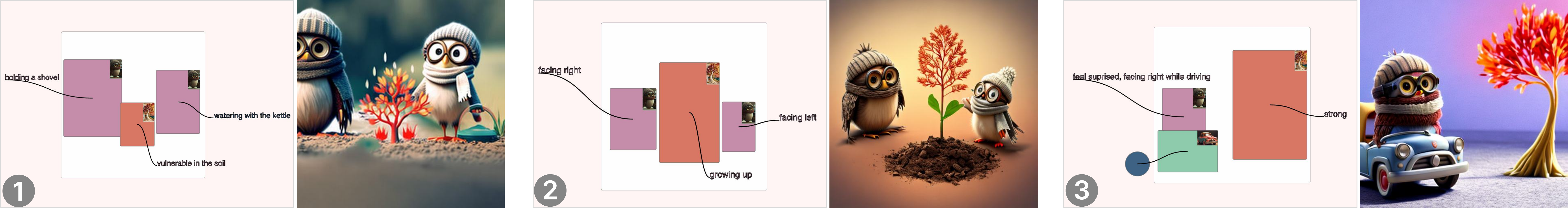}
    \vspace{-6mm}
    \caption{P3 generated results with \sysname{}, illustrating a story where (1) two owls plant a tree together, (2) nurture it with care, and (3) later, one owl drives by and happily witnesses the tree's growth into tall and strong.}
    \label{fig:study2_results_p3}
    \Description{P3 generated results, illustrating a story where (1) two owls plant a tree together, (2) nurture it with care, and (3) later, one owl drives by and happily witnesses the tree's growth into tall and strong.}
\end{figure*}

\begin{figure*}[tbp]
    \centering
    \includegraphics[width=\linewidth]{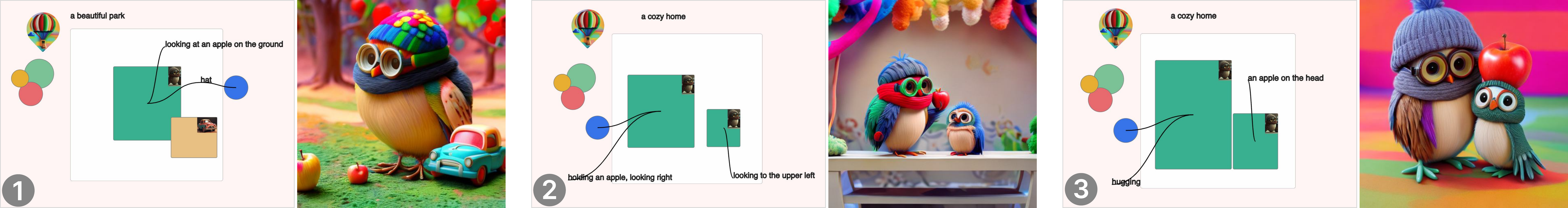}
    \vspace{-6mm}
    \caption{P6 generated results with \sysname{}. The user intends to describe (1) an owl discovers some apples in a park, picks one, and (2) brings it home to share with a friend, and (3) his friend puts the apple on the head, sharing a happy moment together.}
    \label{fig:study2_results_p6}
    \Description{P6 generated results. The user intends to describe (1) an owl discovers some apples in a park, picks one, and (2) brings it home to share with a friend, and (3) his friend puts the apple on the head, sharing a happy moment together.}
\end{figure*}

\begin{figure*}[tbp]
    \centering
    \includegraphics[width=\linewidth]{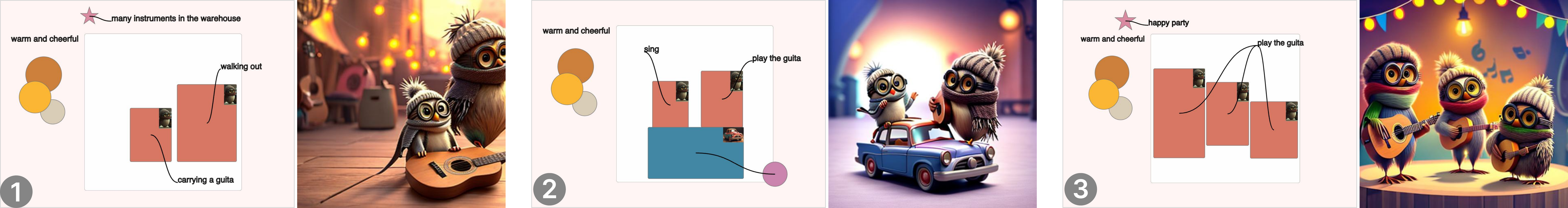}
    \vspace{-6mm}
    \caption{P10 generated results with \sysname{}, depicting a story where (1) the owl and his friend discover and move a guitar from a warehouse, (2) joyfully sing and play the guitar on a car, and (3) invite another friend to join them for a lively music party.}
    \label{fig:study2_results_p10}
    \Description{P10 generated results, depicting a story where (1) the owl and his friend discover and move a guitar from a warehouse, (2) joyfully sing and play the guitar on a car, and (3) invite another friend to join them for a lively music party.}
\end{figure*}

\begin{figure*}[tbp]
    \centering
    \includegraphics[width=\linewidth]{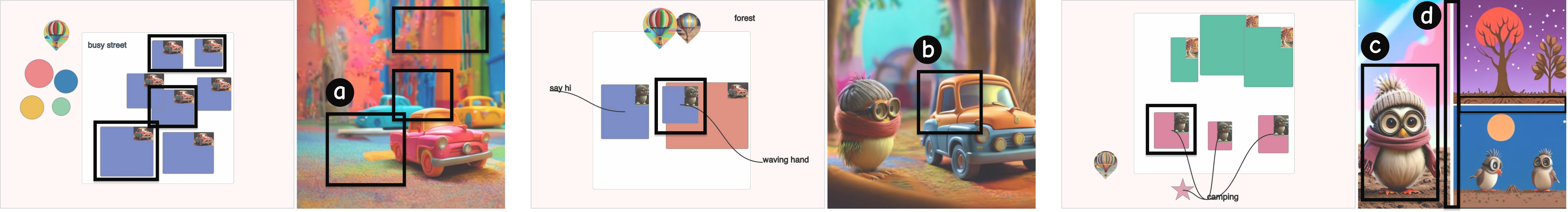}
    \vspace{-6mm}
    \caption{Failure cases: (a) when there are multiple subjects, the model often omits some of them (from P1); (b) when two subjects overlap significantly, the model struggles and incorrectly interprets the interaction as being ``inside'' (from P5); (c) the size of generated subjects sometimes may not align with the user-specified one; and (d) the model occasionally fails to produce a cohesive image, instead generating patchy outputs with visible edges (from P11).}
    \label{fig:failure_cases}
    \Description{Failure cases: (a) when there are multiple subjects, the model often omits some of them (from P1); (b) when two subjects overlap significantly, the model struggles and incorrectly interprets the interaction as being ``inside'' (from P5); (c) the size of generated subjects sometimes may not align with the user-specified one; and (d) the model occasionally fails to produce a cohesive image, instead generating patchy outputs with visible edges (from P11).}
\end{figure*}

\rev{
The factor of \emph{results worth effort} reflects how effectively the model executes users' intent. 
Half of the participants rated it positively, while the rest were less satisfied. 
This disparity arised because, while our pipeline for executing visual lexicons (Section~\ref{sec:executaion_pipeline}) is feasible for simpler cases, its limitations become apparent in complex scenarios, such as handling multiple subjects or intricate relationships, leaving significant room for improvement. 
We showcase some participant-created story using \sysname{}, including planting a tree (Figure~\ref{fig:study2_results_p3}), sharing an apple (Figure~\ref{fig:study2_results_p6}), and hosting a music party (Figure~\ref{fig:study2_results_p10}), where they crafted three-scene narratives effectively. 
However, failure cases (Figure~\ref{fig:failure_cases}) reveal that the model often omits subjects when there are more than three and/or when subjects overlap significantly (\eg, being ``inside'').
Additionally, the model sometimes fails to match specified sizes and occasionally produces patchy images. 
In response, participants typically regenerated outputs with different seeds, which sometimes worked. 
If not, they reduced the complexity, such as removing some subjects or tweaking the layout, to gradually adapt themselves to the model’s capacity limits.
We anticipate that as the base model (currently Stable Diffusion 2.1) continues to grow in size and evolve in architecture, its performance in executing the visual lexicon would improve, thus mitigating this problem.
}

\subsection{Observations and Participants' Feedback}

\subsubsection{Decomposing a reference image into elements is more effective than using it as a whole.}
Participants consistently valued decomposing reference images into individual elements rather than using them as a whole, aligning with their design process. 
P3 noted, \qt{Being able to break down an image into parts lets me mix and match elements in a way that fits my vision, rather than being constrained by the original composition}.
Compared to their previous experiences with tools that only allow for remixing entire images, participants found the ability to recombine multiple decomposed elements and specify their relationships particularly valuable. 
P7 expressed, \qt{With other tools, I'd have to try different images multiple times to get something close to what I want. Here, I can just pull out the pieces I need and arrange them how I like}.

\subsubsection{\textsc{\sysname{}} enhances the sense of control but requires a target in mind.}
Participants unanimously agreed that \textsc{\sysname{}} offers a greater sense of control compared to using natural language prompts. 
They found it particularly useful for defining and manipulating the relationships between elements. 
P4 noted, \qt{it (\textsc{\sysname{}}) makes it so much easier to specify how different parts of my design interact. I feel like I have more direct control over the outcome}.
However, participants also highlighted that this increased control comes with a prerequisite: having a clear idea or target in mind. 
When they were unsure of what they wanted to create, they found that natural language prompts offered a quicker and more flexible starting point. 
As P9 explained, \qt{If I don't have a clear idea, it’s easier to just throw in some random words and see what the AI generates, it's a good way to get inspired}.
These feedback implies a trade-off between control and exploration. 
Natural language allows for broad exploration and can spark new ideas even from vague or random inputs, while \textsc{\sysname{}} excels in depth and precision when users have a rough direction or specific visual properties in mind. 
As P1 put it, \qt{Once I know the general look I'm going for, it (\textsc{\sysname{}}) lets me really arrive there}.

\subsubsection{Direct manipulation on design tokens helps reduce the efforts of refinements}
\textsc{\sysname{}} allows for more precise and enjoyable fine-tuning. 
Participants felt useful to directly manipulate elements without losing the core identity of their design. 
\qt{[With \textsc{\sysname{}},] I can make the adjustments I want without compromising the overall look, I actually very enjoy this refinement process}, shared by P5.
The persistence of design tokens throughout the creative process was particularly valued. 
Participants liked that once a design token was created, it could be reused throughout a session, streamlining the workflow and ensuring consistency across different iterations. 
\qt{I love that I can just drag and drop a copy of a token for reuse,} P8 shared, \qt{It keeps my design consistent without having to start from scratch every time}.
In contrast, participants mentioned the challenges they faced with other tools when trying to make minor refinements. 
In many cases, they found that the visual identity of their work would start to drift with each new text-based prompt, leading to frustration. 
\qt{I’ve tried making refinements in other tools, but often the visuals change too much, even when I just want minor tweaks. It’s so frustrating that I usually just switch to Photoshop to fix it myself}, P12 explained.

\subsubsection{\textsc{\sysname{}} do not require a rigid structure thus enable a flexible thinking flow}
Several (P1, P2, P4, P7, P8, P9, P11) participants mentioned they benefited from freely constructing their visual lexicon without worrying about ordering and format constraints.
They felt \textsc{\sysname{}} can enable them to think non-linearly and creatively.
P7 noted, \qt{with text prompts, I was advised to follow a template for effective prompting, though I don't know if it really matters: start with the overall context, list subjects, describe the view with terms like headshots or close-ups, and then add magic words for styling. It feels like I'm being forced to think in a certain way, which is not how design should work, it is supposed to be messy. You think about this, then that, then come back to adjust one accordingly, the elements are influencing each other}.
P11 expressed frustrations that \qt{different tools often require different prompt formats, making it (prompting with texts) more complicated}. 
In contrast, \textsc{\sysname{}} does not assume a procedural process to construct the visual lexicon, users are free to start from any design aspects and seamlessly navigate between them. 
Overall, the flexibility offered by \textsc{\sysname{}} was seen as a significant advantage, enabling users to engage in a more dynamic and less restrictive creative thinking process.

\subsubsection{Clarity in intent expression help users appropriately understand AI accountability}

Users often face challenges in correctly attributing AI failures due to the lack of transparency in how AI interprets their inputs. 
This can lead to confusion, where users mistakenly believe that failures are due to unclear intent rather than AI's limitations. 
As P5 mentioned, \qt{Sometimes, I'm not sure if the problem is with how I'm phrasing things}.
This uncertainty often results in users repeatedly refining their inputs, leading to unnecessary back-and-forth. 
However, with \textsc{\sysname{}}, users felt more confident in the clarity of their intent expression, such confidence making users can appropriately attribute failures to AI. 
P12 commented that \qt{[With \sysname{},] I know if something goes wrong, it's probably the AI, not me}.

%% file: texfiles/9-discussion.tex
\section{Discussion}
We reflect on the design of \textsc{\sysname{}} and discuss the lessons we learned and the implications for future research.

\subsection{Design Implications}

\subsubsection{Integrating texts into visual-centric paradigm versus embedding visuals into texutal-centric paradigm.}
\textsc{\sysname{}} addresses two key ambiguities in textual prompts: \emph{what} elements to use and \emph{how} to construct them for an intended effect. 
Prior work, like DirectGPT~\cite{masson2023directgpt}, also augments textual prompts by embedding visual symbols to clarify object references, focusing on the \emph{what} aspect. 
However, these two approaches differ fundamentally in their expressiveness of the \emph{how} aspect due to their structural nature.
Visual-centric paradigms like \textsc{\sysname{}} operate in a two-dimensional (2D) space, enabling spatial manipulation and richer exploration of relationships between elements, with text serving as a complement. 
In contrast, text-centric paradigms with embedded visuals, such as DirectGPT, function within a linear, one-dimensional (1D) space, where the narrative sequence is preserved, and visuals enhance object references or enrich textual information.
We argue that the effectiveness of either paradigm is likely task-dependent. 
For visual tasks, the spatial properties of a 2D approach can provide more intuitive and efficient interactions, aiding in design, spatial reasoning, and layout organization. 
However, this may disrupt the narrative flow, making it less suitable for tasks that require step-by-step instructions such as programming. 
Conversely, a text-centric approach with visual enhancements may be better suited for tasks requiring logical reasoning and narrative coherence.
This distinction raises important questions for future research: \emph{when} should one paradigm be chosen over the other? Or is there potential for blending the spatial advantages of visuals with the narrative flow of text to enable a unified paradigm for various tasks?

\subsubsection{Dynamic distributed agency between user and AI: letting users to specify when, where, and how much}
\textsc{\sysname{}} is designed to reduce the ambiguity in design intent expression and thus improve users' sense of control when working with AI.
We recognize that users' required control varied between individuals with different skill levels and varied at different stages in design process.
When users seek to leverage AI's creativity, they often \emph{choose} to leave ambiguity in the visual lexicon for the AI to refine. 
Conversely, when they have a clear vision, they specify their intentions with detailed visual tokens and intricate spatial manipulations, letting AI to execute their vision with precision.
Thus, the agency distribution between user and AI is dynamically changing, also discussed by Satyanarayan~\etal~\cite{Satyanarayan2024Intelligence}. 
In \textsc{\sysname{}}, users can actively and explicitly configure \emph{when}, \emph{where}, and \emph{how much} control they wish to shift to AI through the use of the imagination token.
For instance, \rev{during the Study 2, P2 provided a brief description ``a beautiful park'' as background, and assign a large imagination token, signaling the AI to elaborate. }
Participants in our study valued this flexibility and control.
In contrast, most current text-to-image tools, like DaLLE, MidJourney, and Adobe Firefly, automatically expand and refine users' whole prompts without asking users if they want, leading to unintended results.
This informs the importance of providing an \emph{explicit} way for users to actively delegate control to AI --- managing \emph{when}, \emph{where}, and \emph{how much} --- rather than assuming a fixed agency distribution pre-defined by the system.

\subsubsection{Towards bi-directional visual lexicon construction: enabling both users and AI to be constructors}
Reflecting the current design of \textsc{\sysname{}}, design tokens serve as the communication medium between users and AI, while the constructed visual lexicon acts as a visual abstraction of the generated image. 
However, \textsc{\sysname{}} only allows users to construct the visual lexicon, with AI solely acting as the receiver to execute it.
This workflow presents a challenge: as mentioned by some participants, in the early ideation stage when users may not have a clear vision, they would prefer natural language as a quick starting point. 
But how can users complete the iterative design process without an initial visual lexicon? 
What if they wish to refine an AI-generated image instead? 
The current system of \sysname{} does not fully close this interaction loop.
One potential solution is to enable AI to construct the visual lexicon as well. 
Given an image, AI could automatically extract design tokens and compose a corresponding visual lexicon for users to manipulate. 
This approach is similar to the prior work on \emph{abstraction-driven} color manipulation for image~\cite{shi2024exploring} and motion graphic videos~\cite{shi2024piet}, where the system creates color abstractions for manipulation. 
This informs that visual abstractions can be constructed bi-directionally, with both users and AI acting as constructors.
However, this also raises the questions warrant more in-depth exploration in future work: how to decide the granularity of the AI-generated visual lexicon; what elements should be reified into tokens; and what relationships should be reflected? 

\subsection{Design Opportunities in \textsc{\sysname{}}}

\subsubsection{Diversifying design token types and sources}
Currently, \sysname{} supports limited visual token types including subject, style, and color. 
However, there are more visual elements essential for constructing a successful design, such as camera angle, depth, texture, and material. 
\textsc{\sysname{}} can naturally incorporate them by adding corresponding token types. 
For example, with a camera token, users can specify the camera angle by positioning it in the visual lexicon, which is otherwise hard to express with texts.
It is also possible to automate the token creation process by decomposing an image into low-level design elements~\cite{vinker2023concept}.
\rev{
It is worth noting that \textsc{\sysname{}} currently only supports reference image as the source of visual tokens. 
While starting from references is common, it is certainly not the only approach that designers use. 
Designers could draw inspirations from other sources such as their own memory, using sketches to externalize and articulate their intent~\cite{kirsh2010thinking, jonson2005design}. 
Expanding \textsc{\sysname{}} to incorporate sketch as a source of design tokens is a promising way to accommodate such scenarios.
}

\subsubsection{Customizing and Re-configuring design tokens by users}
While \sysname{} can expand the supported design token types, it is impractical to preset every possible type. 
A valuable future direction is to allow users to make their own tokens. 
Users could define tokens by describing their functions in natural language and providing examples. 
The system would then dynamically generate and support these tokens. 
Furthermore, preset tokens may not always match user's intended usage.
It would be more flexible to allow users to appropriate the pre-designed tokens for their specific needs --- reconfiguring their represented meanings.
For example, while current color tokens only convey proportional information through size, we observed P3 using their position to indicate specific areas on a subject to colorize.
Lastly, it is important to constrain such customization and reconfiguration to remain interpretable by generative AI models to ensure effective interaction. 

\subsubsection{Affording richer manipulations on design tokens} 
\textsc{\sysname{}} currently supports the manipulations of resizing, positioning, grouping, and linking.
However, during the user study, participants attempted additional manipulations beyond those provided. 
For example, in the first task, many participants (7/12) tried to rotate the token to indicate the pumpkin's position. 
Similarly, some (3/12) wanted to move subject tokens forward or backward to specify spatial relationships.
These observations reveal the potential for our interaction paradigm to support more intricate user intentions. 
Beyond spatial relationships, other behaviors like blending could also be supported.
While richer manipulations can enhance user experience, they also introduce complexity. 
Ideally, users should rely on \emph{technical reasoning} --- intuitively understanding how manipulations affect outcomes --- rather than \emph{procedure learning} --- memorizing steps to achieve desired effects~\cite{beaudouin2021generative}.
Future work will investigate what manipulations are both desired and natural for users to express their intentions effectively.

\subsubsection{Propagating the modifications of the design token}
A limitation of the current version of \sysname{} is that only textual tokens are editable, while visual tokens, once created, cannot be re-linked to another visual element. 
For example, if a user creates a subject token of an owl and uses it in multiple designs, changing this owl to a rabbit requires reconstructing everything from scratch.
To address this, a possible improvement is to allow design token modification with automatic propagation to all instances. 
Since we designed the visual tokens to be persistent and every time users construct a visual lexicon, they drag a copy of the original token, this creates a natural link between the original and its copies.
Leveraging this link, once users modify the original token, such as re-attaching it to a new subject in another image, we can propagate this modification to all its copies in related visual lexicons. 
Such a propagation mechanism can allow designers to quickly compare different visual candidates and streamline their workflow.

\subsubsection{Beyond static graphic design: extending \textsc{\sysname{}} to video, 3D scene, and other co-creation tasks with AI}
While \textsc{\sysname{}} is initially designed for static graphic design, its visual-centric interaction paradigm holds significant potential for broader applications in general AI-assisted visual design tasks. 
In video creation, for instance, \textsc{\sysname{}} could adapt the visual lexicon to a timeline-based structure, where individual lexicons construct each scene.
Similarly, in 3D scene modeling, where spatial relationships are more complex, \textsc{\sysname{}} could extend 2D design tokens into 3D tokens and extend the 2D manipulations to 3D operations. 
While token design and manipulations may be domain-specific, the fundamental interaction logic of using direct manipulation on tokens to construct elemental relationships remains coherent and consistent across different design domains.
We believe \textsc{\sysname{}} opens up the possibility to offer designers a unifying design language to communicate with AI across the broad creative landscape.

\subsection{Limitations}

\subsubsection{\rev{\textsc{\sysname{}} might fail in describing unseen visuals beyond recombination.}}
\rev{
Each modality has special strengths and weaknesses in its ability to communicate particular concepts~\cite{buxton1989natural}.
While \textsc{\sysname{}} excels at referring to elements and describing spatial relationships, its reliance on existing visuals might lead to design fixation~\cite{jonson2005design}, where designers may unconsciously adhere to what has already been known or available.
In contrast, natural language could describe unseen visuals that go beyond recombination (\eg, \qt{a cute sock with a human-like face}) or ideas that might seem unreasonable (\eg, \qt{time is melting}). 
Therefore, our proposed visual-centric interaction paradigm is not intended to replace the text-centric approach.
Instead, it is important for designers to strategically choose the most suitable modality based on the design context and the level of originality they seek to achieve.
}

\subsubsection{Visual lexicon extraction could be improved.}
Our implementation of visual lexicon execution relies heavily on off-the-shelf computer vision techniques.
Despite that we selected the state-of-the-art ones at the time of developing \sysname{}, these techniques have inherent limitations that impact our system's capacity. 
Currently, \sysname{} supports only 4-6 subject tokens. 
This constraint arises because we use the \texttt{Break-A-Scene}~\cite{avrahami2023break} approach to preserve each subject's visual identity. 
However, our experiments show that when there are more than six subjects, the performance in maintaining visual identity drops significantly.
The scope of this work is not to improve the performances of computer vision models, but we do hope this work informs the importance of computer vision research to push the progress forward --- training the model to be aware and preserve visual details rather than solely taking natural language as input.

\subsubsection{Inference and computation costs could hinder user experience.}
Due to the high computational costs and inference time of diffusion models, we cannot support on-the-fly inference and immediate feedback. 
Our visual lexicon execution pipeline sequentially handles different design aspects (layout, style, and color), each requiring a 50-step diffusion model inference (around 30 seconds).
As a result, users must click a button to generate and wait for the results, interrupting their design experience to some degree.
We envision that as inference time and computation costs decrease, users will no longer need to click the generation button after constructing the visual lexicon. 
Instead, they will receive immediate feedback while manipulating the tokens. 
This will enable users to instantly see the effects of their actions, providing a smoother co-creation experience with generative models.

\subsubsection{Study results might not be generalizable for design novices.}
Participants in our study are experienced designers, all with a minimum of 3 years of design experience. 
These participants are trained to approach design problems visually, and \textsc{\sysname{}} was specifically designed to align with this visual-centric mental model. 
As a result, the study’s findings may not generalize to novice designers or casual users who lack this level of expertise.
It is uncertain whether novice users could adapt to this visual-centric paradigm and fully leverage the fine-grained control.


%% file: texfiles/10-conclusion.tex
\section{Conclusion}

In this paper, we introduce \textsc{\sysname{}}, a \emph{visual-centric} interaction paradigm that allows users to express design intent more effectively. \textsc{\sysname{}} reifies primitive design elements from reference images into \emph{interactive}, \emph{reusable} design tokens, enabling users to specify \emph{what} elements to use and \emph{how} to construct them towards the desired effect. 
We implement \sysname{} to exemplify how state-of-the-art AI models can execute users' intent expressed through \textsc{\sysname{}}.
In a user study, experienced designers found it easier to describe visual details and relationships with fewer mental demands through \textsc{\sysname{}}. 
They efficiently explored design alternatives by reusing tokens and performed refinements more quickly, particularly for complex designs. 
Designers preferred \textsc{\sysname{}} over textual-centric prompting approach, valuing the sense of control it provided when they had design ideas in mind.
Moving forward, we plan to extend \textsc{\sysname{}} to include more element types and operations, broadening its expressive capabilities. 
The design implications derived from this work shed light on future research to design effective interaction mediums for human-AI co-creation.

%% file: texfiles/12-table.tex
\begin{table*}[h]
\caption{The table records participants' demographic information, including gender, age, occupation, and experiences of graphic design in years (Design Exp.), self-rated frequency of Generative AI usage (GenAI Freq.), and GenAI tools they frequently use.}
\vspace{-3mm}
\label{tab:evaluation_demographic}
\begin{tabular}{lcclccl}
\toprule
\textbf{ID} & \textbf{Gender} & \textbf{Age}   & \textbf{Occupation}  & \textbf{Design Exp.} & \textbf{GenAI Freq.} & \textbf{GenAI Tools} \\
\midrule
P1 & F & 25 & Design New Grad & 5 & 3 &  Midjourney \\
P2 & M & 27 & Design Researcher & 5 & 5 &  Dall-E  \\
P3 & F & Not Reveal & Designer & 4 & 3 & Midjourney   \\
P4 & M & 28 & Visual Designer & 5 & 3 &  Midjourney  \\
P5 & F & 25 & 3D Artist & 5 & 5 &  Dall-E, Midjourney \\
P6 & F & 26 & Exhibition Designer & 4 & 3 &  MidJourney  \\
P7 & M & 27 & Graphic Designer & 5 & 2 &  MidJourney  \\
P8 & F & 26 & UX Designer & 5 & 4 &  MidJourney  \\
P9 & M & 25 & Technical Aritist & 3 & 3 &  MidJourney  \\
P10 & M & 25 & Architect & 3 & 2 &  Dall-E  \\
P11 & M & 25 & Graphic Designer & 5 & 4 &  MidJourney \\
P12 & M & 30 & Visual Designer & 5 & 3 & MidJourney  \\
\bottomrule
\end{tabular}
\end{table*}

\begin{table*}[h]
\caption{Rating rubric for the quality of participants' expressions in terms of the element coverage, clarity of size, position, style, and color. Raters were rated on a 7-point Likert Scale where 1 means very low and 7 means very high.}
\label{tab:rubric}
\begin{tabular}{lcll}
\toprule
\textbf{Item}                                                  & \textbf{Score} & \textbf{Criteria} &    \\
\midrule
\multicolumn{1}{c}{\multirow{3}{*}{Element Coverage}} & 7     &  All key elements from the target image are included and accurately represented.      &    \\
\multicolumn{1}{c}{}                                  & 4     &   Three elements are missing.       &    \\
\multicolumn{1}{c}{}                                  & 1     &   Six or more elements are missing.      &    \\
\midrule
\multirow{3}{*}{Clarity of Size}                      & 7     &  The relative size of all elements is clearly and accurately described.        &    \\
                                                      & 4     &  The relative size of around half elements is described, but some ambiguity exists.       &   \\
                                                      & 1     &  The size of elements is highly unclear or not described.       &  \\
\midrule                                                      
\multirow{4}{*}{Clarity of Position}                  & 7     &  The position of all elements is clearly and accurately described relative to each other.        &    \\
                                                      & 4     &  The position of around half elements is described, but some spatial relationships \\ & & are ambiguous.        &    \\
                                                      & 1     &  The position of elements is highly unclear or not described.        &    \\
\midrule
\multirow{5}{*}{Clarity of Style}                     & 7     &  The global style is clearly and accurately described. \\  & & 1) Referred to the style of the first and/or the second reference images; or  \\ & & 2) style descriptors such as  “minimal/simplicity/geometric/abstract” or similar ones.         &    \\
                                                      & 4     &  The global style is described, but it is not apparent how it relates to the target image.        &    \\
                                                      & 1     &  The style of elements is unclear or not described.        &    \\
\midrule
\multirow{4}{*}{Clarity of Color}                     & 7     &  The color of all elements is clearly and accurately described.        &    \\
                                                      & 4     &  The color of about half elements is described and close to the target colors, \\ & & but the rest are ambiguous or missing.        &    \\
                                                      & 1     &  The color of elements mostly described far away from the target image or not described.        &    \\
\bottomrule                                            
\end{tabular}
\end{table*}

%% file: texfiles/11-appendix.tex
\section{Appendix}

\subsection{Implementation Details}
The front-end user interface of \sysname{} was built using React.js as the primary framework.
Most of the design token management functionalities, such as the creation, deletion, and manipulation, were implemented using D3.js. 
The rest of the interface components, such as the buttons and icons, were taken from the Material UI library and customized to fit the needs of the application. 
The server-side rendering for API calls is handled by fastAPI.
The back-end model fine-tuning and inferences for visual lexicon execution are written in Python and performed on an 80G A100 GPU.

\subsection{Participants' Demographic Information}
\label{sec:demographic}
Table~\ref{tab:evaluation_demographic} describes the detailed demographic information of participants in our user study.

\subsection{Rating Rubric for Expressions in Study 1}
\label{sec:rubric}

Table~\ref{tab:rubric} lists the detailed rating rubric for the external scorers in Study 1 to rate participants' expressed intention. 
